\documentclass[prb,superscriptaddress,twocolumn, floatfix]{revtex4-2}
\usepackage{booktabs}
\usepackage{makecell}
\usepackage{mathtools}
\usepackage{verbatim}
\usepackage{graphicx}
\usepackage{xcolor}
\usepackage{amsmath}
\usepackage{amsfonts}
\usepackage{amsthm}
\usepackage{amssymb}
\usepackage{pifont}
\usepackage{amsbsy}
\usepackage{wasysym}
\usepackage{bm}
\usepackage{mathrsfs}
\usepackage{color}
\usepackage{hyperref}
\usepackage{soul}
\usepackage{multirow}
\usepackage{bbm}
\usepackage{physics}
\hypersetup{
    colorlinks=true,
    linkcolor=red,        
    citecolor=blue,
    breaklinks=true,
    urlcolor=blue
}
\DeclareMathAlphabet{\mathcal}{OMS}{zplm}{m}{n}

\newcommand{\sn}{\mathrm{sn}}
\newcommand{\cn}{\mathrm{cn}}
\newcommand{\dn}{\mathrm{dn}}

\newcommand{\opS}[2]{\hat{S}_{#2}^{#1}}
\newcommand{\ops}[2]{\hat{s}_{#2}^{#1}}

\usepackage{orcidlink}

\begin{document}

\title{Stabilization of Granovskii–Zhedanov Scars of the XYZ Quantum Spin Chain via Non-Hermitian Spin Relaxation}

\author{Dhiman Bhowmick}
\affiliation{Department of Physics, National University of Singapore, Singapore 117551}

\date{\today}

\begin{abstract}

The Granovskii–Zhedanov (GZ) states are exact scar states of the spin-$S$ XYZ chain for $S\geq 1$.  
As a result, local quantum information encoded in a GZ state remains preserved under the unitary dynamics of the XYZ Hamiltonian; thus, these states evade thermalization and violate ergodicity despite the system being otherwise nonintegrable and chaotic. 
However, in realistic experimental settings, the realization of an ideal XYZ Hamiltonian is not possible, as perturbations are inevitable.
These perturbations ultimately lead to the decay and thermalization of the GZ state.
We study the stability and dynamics of GZ states in the presence of generic perturbations and propose physically realistic mechanisms to stabilize them.
We show that the product structure of the GZ state allows its lifetime to be enhanced in the presence of an external helical magnetic field, which slows down thermalization but does not prevent it at long times.
We further demonstrate that the inclusion of effective non-Hermitian spin relaxation processes can substantially stabilize the GZ states, leading to a nonequilibrium steady state with finite fidelity with GZ state.
Such dissipative processes can naturally originate from mechanisms such as Purcell-enhanced spontaneous emission or spin–lattice relaxation in the presence of the helical magnetic field.
Using infinite time-evolving block decimation (iTEBD) and exact time evolution, we systematically analyze the dynamics and robustness of the GZ states in the perturbed non-Hermitian XYZ model. 
To connect with experimental platforms, we introduce a Hubbard model that maps onto the XYZ spin system and propose that ring-shaped optical lattices may provide a viable route for realizing and stabilizing GZ states.
Finally, we present an equivalent Lindblad description of the effective non-Hermitian dynamics.

\end{abstract}

\maketitle


\section{Introduction}


The study of quantum thermalization has emerged as a vibrant field, driven in large part by its implications for understanding how isolated quantum systems approach equilibrium\,\cite{ETH1,ETH2,ETH3,ETH4}. 
Central to this framework is the Eigenstate Thermalization Hypothesis (ETH), which posits that, in generic nonintegrable many-body systems, highly excited states in the middle of the energy spectrum exhibit thermal properties consistent with statistical mechanics\,\cite{ETH_more1, ETH_more2, ETH_more3,ETH_more4,ETH_more5}.
However, this paradigm is not universally valid. It is strongly violated in systems such as integrable models and those exhibiting many-body localization, where an extensive number of conserved quantities or strong disorder prevents thermalization\,\cite{MBL1,MBL2}.
A particularly intriguing deviation from ETH arises in the form of quantum many-body scars (QMBS), which constitute a weak breakdown of ergodicity. 
Unlike fully localized or integrable systems, QMBS occur in nonintegrable systems and involve only a small subset of nonthermal eigenstates embedded within a thermal spectrum, thereby enabling atypical dynamical behavior while the majority of states remain thermal.
Therefore, breaking the ETH weakly.
The current knowledge of this phenomena ranges in a variety of interesting systems, including the PXP chain\,\cite{PXP1,PXP2}, the 1D transverse field Ising model\,\cite{Ising1,Ising2}, the fermionic Hubbard model\,\cite{Hubbard1, Hubbard2, Hubbard3, Hubbard4}, quantum Hall systems\,\cite{Hall1, Hall2}, fracton topological ordered systems\,\cite{Fracton1, Fracton2}, AKLT spin chain\,\cite{AKLT1, AKLT2, AKLT3, AKLT4}, the spin-1 XY model\,\cite{XY1, XY2}, frustrated spin systems\,\cite{FrustratedSpin1, FrustratedSpin2} and more\,\cite{Review1, Review2}.
Experimental observations of such phenomena range from ultracold atoms to superconducting circuits.
Owing to their dynamical stability, scar states have been proposed as promising candidates for applications in quantum sensing and for probing disorder in quantum many-body systems\,\cite{DisorderDetection_QSimulator_SPXP, QuantumSensing}.

The spin system, in particular, has been a strong focus in studying many-body phenomena.
Particularly, for the XYZ spinchain product helical scar states exist, originally introduced by Granovskii and Zhedanov, this scar state is hereafter referred to as the Granovskii–Zhedanov (GZ) scar\,\cite{GranovskiiZhedanov1, GranovskiiZhedanov2}.
Such states has been rediscovered again very recently in various theoretical and experimental studies\,\cite{AssyemtricDecay,ExactSpinHelix,GZ_ModernAlgebra, BethePhantom1, BethePhantom2, BethePhantom3,BethePhantom5}.
For spin values $S\geq 1$, the GZ states are the exact QMBS of XYZ spin chains and thus preserve the local quantum information intact.
However, realizing an ideal XYZ Hamiltonian in experiments is not viable, and thus, in the realistic experimental systems, these states will eventually thermalize in the presence of perturbations.
Although scar states typically exhibit slow thermalization in the presence of perturbations\,\cite{SlowThermalization}, the development of stabilization protocols is essential for preserving quantum information in such states.
Previous studies have attempted to control many-body dynamics or enhance the lifetime of scar states using specific driving protocols\,\cite{DrivingProtocol1,DrivingProtocol2, DrivingProtocol3, DrivingProtocol4,DrivingProtocol5}.
Alternatively, approaches based on open quantum systems—either in non-Hermitian or Lindbladian frameworks—have proven effective\,\cite{Stabilize2,Stabilize3}.
In particular, stabilization of spin-helix states in the spin-$1/2$ XXZ model using the non-Hermitian terms has been explored in Refs.\,\cite{HelixDynamics1, SGA2, VladislavPopkov1, VladislavPopkov2}.
In this work, we demonstrate a controlled enhancement of the lifetime of GZ states using experimentally realistic physical mechanisms.
Owing to their product structure, the application of a helical magnetic field that mirrors the spin configuration of the GZ state significantly enhances its stability.
While this field strongly suppresses thermalization, the system nevertheless thermalizes at long times even in its presence.
Specifically, we demonstrate that, on top of the helical magnetic field, effective non-Hermitian spin relaxation processes naturally emerge in a spin system coupled to a bath. 
These processes drive the system to a non-equilibrium steady state with finite fidelity with the GZ state, thereby stabilizing it.
We identify that such dissipative mechanisms can naturally arise from experimentally relevant processes, including Purcell-enhanced spontaneous emission and spin–lattice relaxation.
Using iTEBD simulations and exact diagonalization techniques, we systematically investigate the dynamics and stabilization of GZ scar states in the non-Hermitian XYZ spin system in the presence of various perturbations and analyze the resulting behavior.
Motivated by potential experimental relevance, we introduce a Hubbard model that maps onto the XYZ spin system and suggest that ring-shaped optical lattices may provide a promising platform to realize and stabilize GZ scar states.
Our work provides a concrete route toward engineering a robust scar state in a spin system.


\section{Theoretical Framework}
\label{Sec::Theoretical Framework}
\begin{figure}[tb]
\centering
\includegraphics[width=0.45\textwidth]{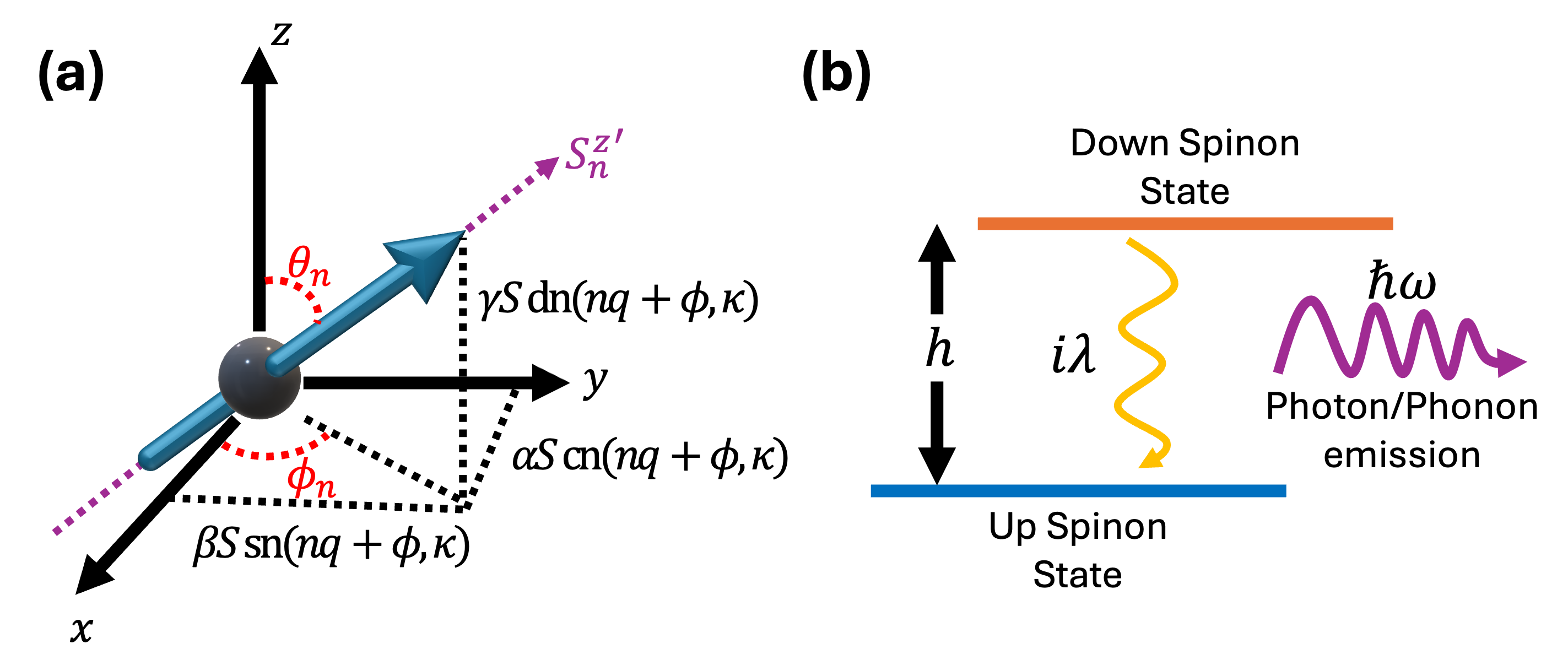}
\caption{
(a) Direction of the spin (denoted by the blue arrow) at the $n$-th site in a GZ state.
This schematic illustrates the parameters $\theta_n$ and $\phi_n$ appearing in Eq.\,\ref{Eq::GZScar}, together with the corresponding expectation value of the components of the local spin operator $ \boldsymbol{\hat{S}}_n$.
It also indicates the $z^\prime$-axis, which is aligned with the local spin direction, such that the spin component along this axis is denoted by $S_n^{z^\prime}$.
(b) Schematic of the spin relaxation process. 
In the presence of an external magnetic field $|\boldsymbol{h}_n|=h$, the local degenerate up and down spinon levels split and are separated by an energy gap $h$.
The non-Hermitian term proportional to $i\lambda$ physically represents spinon decay from the down spinon state to the up spinon state.
This relaxation process is accompanied by the emission of a photon or a phonon, depending on the nature of the environmental bath available.
}
\label{fig::GZ_Schematic}
\end{figure}

In this section, we introduce the Granovskii–Zhedanov (GZ) state for the XYZ spin Hamiltonian and discuss the associated non-Hermitian spin relaxation mechanism.
The GZ state is an extraordinary product eigenstate of the XYZ spin exchange system, which exists for any spin $S$\,\cite{GranovskiiZhedanov1, GranovskiiZhedanov2}.
It was introduced for the one-dimensional XYZ spin chain, but it can also exist in two or higher dimensional lattices\,\cite{GZ_ModernAlgebra, ExactSpinHelix, FelixGraphicalConstruction}.
The one-dimensional XYZ Hamiltonian is given by,
\begin{equation}
    \hat{H}_0=\sum_{n=1}^N \left[
    J_x\opS{x}{n}\opS{x}{n+1}
    +J_y\opS{y}{n}\opS{y}{n+1}
    +J_z\opS{z}{n}\opS{z}{n+1}
    \right],
    \label{Eq::XYZ_Hamiltonian}
\end{equation}
where $\opS{\alpha}{n}$ are the spin-$S$ operators.
A periodic boundary condition is imposed by identifying site $(N+1)$ with site $1$, or alternatively, the infinite-system limit ($N \to \infty$) is considered.
We can determine the paramters $q$ and $\kappa$ from the Hamiltonian coefficients, as follows,
\begin{equation}
    \dn\,(q,\kappa)=\frac{J_x}{J_y}, \quad
    \cn\,(q,\kappa)=\frac{J_z}{J_y}, \quad
    \kappa=\frac{J_y^2-J_x^2}{J_y^2-J_z^2},
\end{equation}
where $\cn$ and $\dn$ are the Jacobi elliptic functions, whereas $\kappa$ is known as the elliptic modulus. 
These parameters, along with a system-independent parameters  $-1\leq\gamma\leq 1$ and $0\leq\phi<2\pi$ determine the GZ state as follows,
\begin{gather}
    \ket{\Psi_{\text{GZ}}}=\prod_n \exp(- i \opS{z}{n} \phi_n) \exp(-i \opS{y}{n} \theta_n) \,\ket{\uparrow_n},    
    \notag \\
    \text{with, }
    \theta_n=\cos^{-1}\left(\frac{\langle{\opS{z}{n}}\rangle}{S}\right), \;\;
    \phi_n=\text{atan}2\left(\langle{\opS{y}{n}}\rangle,\langle{\opS{x}{n}}\rangle\right),
    \notag \\
    \text{where,}
    \notag\\
    \left\langle \boldsymbol{\hat{S}}_n\right\rangle
    =S\Big(
    \alpha\,\cn(nq+\phi,\kappa),\,
    \beta\,\sn(nq+\phi,\kappa),\,
    \gamma\,\dn(nq+\phi,\kappa)
    \Big),
    \notag \\
    \alpha=\sqrt{1-\gamma^2},\, 
    \beta=\sqrt{1-\gamma^2+\kappa^2\gamma^2}.
    \label{Eq::GZScar}
\end{gather}
Here, $\ket{\uparrow_n}$ represents the local spin state at site n corresponding to the highest eigenvalue of $\hat{S}_n^z$.
The GZ state can be pictorially understood by examining the expectation values of the local spin operators, $\left\langle \boldsymbol{\hat{S}}_n\right\rangle$, which specify the orientation of the spin at each lattice site $n$.
This is pictorially demonstrated in Fig.\,\ref{fig::GZ_Schematic}.
Several additional remarks on the GZ state are in order.
First, Eq.\,\ref{Eq::GZScar} does not correspond to a unique state, but instead defines an infinite family of states. 
Varying the free parameters $\gamma$ and $\phi$ enables access to different GZ states within the $4NS$-dimensional GZ scar subspace\,\cite{GZ_ModernAlgebra}. 
For simplicity, and without loss of generality, we set the variable $\phi$ to zero throughout this study.
Second, the GZ state can be regarded as a generalization of the helical state of the easy-plane one-dimensional XXZ spin chain to the XYZ model.
Finally, for a finite periodic chain of length $N$, the allowed GZ scar states are restricted to $q$-values (or wave vectors) of the form  $4pK(\kappa)/N$, where $p$ is an integer and $K(\kappa)$ denotes the complete elliptic integral of the first kind.
 A proof that the GZ states are eigenstates of the XYZ Hamiltonian is given in Appendix.\,\ref{Appendix::Sec::GZ_EigenState}.
We also provide explicit expressions for these states for several spin values in the local $S^z$-basis in the same Appendix.

The structure of the GZ states suggests that the application of a helical magnetic field of the following form is one of the straightforward ways to stabilize such state,
\begin{equation}
\begin{split}
    &\hat{H}_h=-h\sum_{n=1}^N \boldsymbol{B}_n \cdot \boldsymbol{\hat{S}}_n,
    \\
    &\boldsymbol{B}_n=\left(
        \alpha\, \cn\,(nq,\kappa),\,
        \beta\, \sn\,(nq,\kappa),\,
        \gamma\, \dn\,(nq,\kappa)
    \right),
    \label{Eq::HelicalMagneticField}
\end{split}
\end{equation}
where the magnetic field $\boldsymbol{\hat{B}}_n$ is aligned along the local spin directions of the GZ state (see Fig.\,\ref{fig::GZ_Schematic}).
In addition to the helical magnetic field, coupling to a suitable environmental bath gives rise to a local spin-relaxation process (see Section.\,\ref{Sec::Physical Realization}).
Such a local spin relaxation process can be described by the following non-Hermitian term,
\begin{equation}
    \hat{H}_{\lambda}
    =
    \sum_{n=1}^N i\lambda\opS{z^\prime}{n},
    \label{Eq::non-Hermitian1}
\end{equation}
where the primed operator is defined in a local coordinate system in which the $z$-axis (denoted as the $z^\prime$-axis) is aligned with the spin direction at site $n$. 
The corresponding spin component along this axis is denoted by $S_n^{z^\prime}$ (see Fig.\,\ref{fig::GZ_Schematic}(a)).
The identification of this non-Hermitian term as a spin-relaxation process can be understood within the Schwinger boson framework, which is defined as,
\begin{equation}
\begin{split}
    &\opS{+^\prime}{n}=\hat{c}_{n\uparrow}^\dagger \hat{c}_{n\downarrow},\,
    \opS{-^\prime}{n}=\hat{c}_{n\downarrow}^\dagger \hat{c}_{n\uparrow},\,
    \\
    &\opS{z^\prime}{n}=\frac{1}{2}\left(\hat{c}_{n\uparrow}^\dagger \hat{c}_{n\uparrow}-\hat{c}_{n\downarrow}^\dagger \hat{c}_{n\downarrow}\right),\,
    \\
    &\text{with constraint, } 
    \hat{c}_{n\uparrow}^\dagger \hat{c}_{n\uparrow}+\hat{c}_{n\downarrow}^\dagger \hat{c}_{n\downarrow}=2S.
\end{split}
\end{equation}
The creation and annihilation operators $\hat{c}_{n\uparrow}^\dagger$, $\hat{c}_{n\downarrow}^\dagger$, $\hat{c}_{n\uparrow}$ and $\hat{c}_{n\downarrow}$ correspond to Schwinger bosons.
Alternatively, the particles associated with the creation operators $\hat{c}_{n\uparrow}^\dagger$ and $\hat{c}_{n\downarrow}^\dagger$ are also known as up and down spinons, respectively.
The GZ state can be interpreted as the vacuum state made of the up-spinons $\hat{c}_{n\uparrow}^\dagger$ and the non-Hermitian term explicitly in terms of the spinons, 
\begin{equation}
    \hat{H}_{\lambda}
    =
    \frac{i\lambda}{2}
     \sum_{n=1}^N (\hat{c}_{n,\uparrow}^\dagger \hat{c}_{n,\uparrow}
     -
     \hat{c}_{n,\downarrow}^\dagger \hat{c}_{n,\downarrow}).
    \label{Eq::non-Hermitian2}
\end{equation} 
Physically, a positive $\lambda$ characterizes the decay rate of down spinons and simultaneously the gain rate of up spinons.
Therefore, the non-Hermitian term represents an onsite spin-relaxation process, which helps to stabilize the GZ state.
This spin relaxation, as a down spinon decay and up spinon gain, is illustrated schematically in Fig.\,\ref{fig::GZ_Schematic}(b).
The non-Hermitian term in Eq.\,\ref{Eq::non-Hermitian1} is further represented in terms of spin operators in the original frame of reference in the following manner,
\begin{equation}
    \hat{H}_\lambda 
    =i \lambda \sum_{n=0}^{N} 
    \boldsymbol{B}_n\cdot\boldsymbol{\hat{S}}_n\,.
\label{Eq::non-Hermitian2}
\end{equation}
Mathematically, a non-Hermitian term for stabilizing the GZ state is not unique: any operator for which the GZ state is an eigenstate can be used to construct a non-Hermitian stabilizer term (see Appendix.\,\ref{Appendix::Sec::AlternativeStabilizers}).
Previous studies have also introduced various non-Hermitian mathematical constructs to stabilize the GZ state for the spin-$1/2$ XXZ systems\,\cite{HelixDynamics1, SGA2}.
In contrast, we emphasize that our approach is physically motivated, with the non-Hermitian term representing an onsite spin-relaxation process that naturally arises in systems due to the Purcell-enhanced spontaneous decay or onsite magnetostriction (see Section.\,\ref{Sec::Physical Realization}).

\begin{widetext}

\begin{figure}[tb]
\centering
\includegraphics[width=\textwidth]{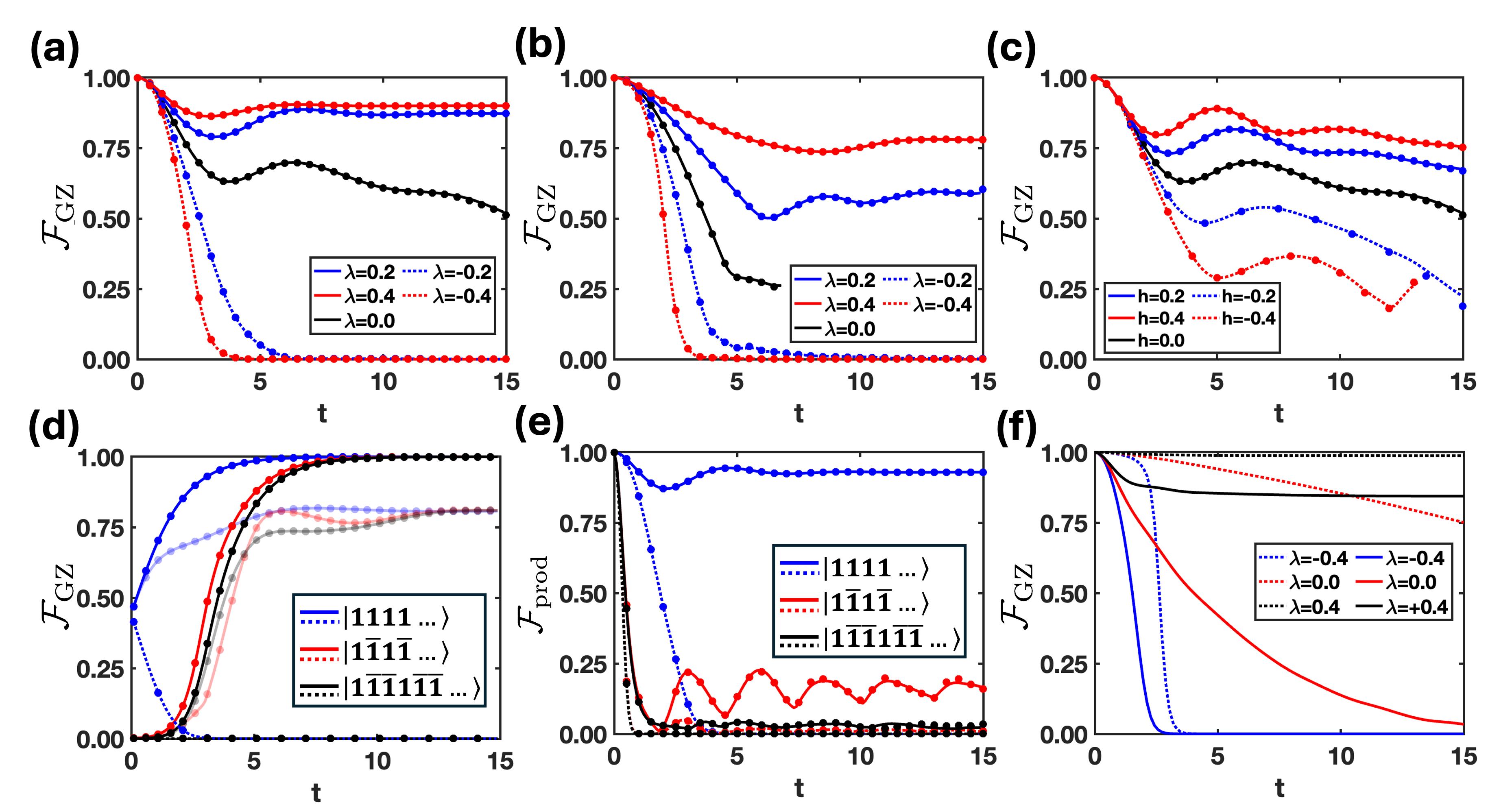}
\caption{
Numerical plots showing the fidelity as a function of time for different scenarios.
In all the simulations, the system is initiated with the GZ state except for the figures (d) and (e).
Additionally,  in all the simulations, the fidelity of the time-evolved state is measured relative to the GZ state, except for the figure (e).
(a) Fidelity dynamics under the Hamiltonian in Eq.\,\ref{Eq::Hamiltonian} with onsite perturbation $\mu = 0.3$, shown for different values of the non-Hermitian strength $\lambda$, with the system initialized in the GZ state.
(b)
Fidelity dynamics under the Hamiltonian in Eq.\,\ref{Eq::Hamiltonian} with hopping perturbation $\mathfrak{t}=0.3$ (and zero onsite perturbation), shown for different values of the non-Hermitian strength $\lambda$, with the system initialized in the GZ state.
(c) Fidelity dynamics under the Hamiltonian in Eq.\,\ref{Eq::Hamiltonian} in the presence of a helical magnetic field, shown for different values of the field strength $h$ and an onsite perturbation $\mu = 0.3$, with the system initialized in the GZ state.
(d) Generating GZ state starting from the several simple product states, namely $\ket{1111\ldots}$, $\ket{1\bar{1}1\bar{1}\ldots}$, $\ket{1\bar{1}\bar{1}1\bar{1}\bar{1}\ldots}$, under the Hamiltonian in Eq.\,\ref{Eq::Hamiltonian} in the presence of non-Hermitian strengths $\lambda=0.4$ (solid lines) and $\lambda=-0.4$ (dashed lines).
Dense [transparent] curves correspond to dynamics in the absence ($\mu=0.0$) [presence ($\mu=0.3$)] of the onsite perturbation.
The parameter $q$ for the Hamiltonian and the GZ state is fixed to $q=4K(\kappa)/6$.
(e)
Fidelity dynamics of the non-eigen product states $\ket{1111\ldots}$, $\ket{1\bar{1}1\bar{1}\ldots}$, and $\ket{1\bar{1}\bar{1}1\bar{1}\bar{1}\ldots}$ under the XYZ Hamiltonian in the presence of the corresponding non-Hermitian spin-relaxation term in Eq.\,\ref{Eq::NonHermitianNonEigenState}, shown for $\lambda = 0.4$ (solid lines) and $\lambda = -0.4$ (dashed lines), with an onsite perturbation $\mu = 0.3$. 
The system is initialized in the respective product state.
We observe that stabilization of the ferromagnetic state $\ket{1111\ldots}$ is achievable, owing to its large overlap with the GZ scar subspace. In contrast, the non-eigen product states $\ket{1\bar{1}1\bar{1}\ldots}$ and $\ket{1\bar{1}\bar{1}1\bar{1}\bar{1}\ldots}$ cannot be stabilized.
(f) Fidelity dynamics for a finite-size system with $N = 9$ and periodic boundary conditions under the Hamiltonian in Eq.\,\ref{Eq::Hamiltonian}, in the presence of local perturbations $\mu_1 \opS{z}{1}$ (dashed lines, $\mu_1=0.3$) and $\mu_2 \opS{+}{1}\opS{-}{1}\opS{-}{2}\opS{+}{2}$ (solid lines, $\mu_2=0.3$), shown for different values of the non-Hermitian strength $\lambda$ and with an onsite perturbation $\mu = 0.3$.
The simulations are done for parameters $\kappa=0.3$, $\gamma=0.5$, $q=4K(\kappa)/3$.
Unless stated otherwise, the default parameters for the Hamiltonian and the GZ state are set to $\kappa = 0.5$, $\gamma = 0.8$, $q = 4K(\kappa)/5$, $S = 1$, $h=0.0$, $\mu=0.3$, and $\mathfrak{t}=0.0$.
To ensure numerical convergence, we use two different iTEBD parameter sets, $(\delta\tau = 10^{-3}, \chi = 120)$ and $(\delta\tau = 5 \times 10^{-4}, \chi = 140)$, represented by dots and solid (or dashed) lines, respectively, in (a)–(e).
}
\label{fig::Dynamics}
\end{figure}

\end{widetext}

\section{Stability Against Perturbations}
\label{Sec::Stability}

To investigate the stabilization of the GZ state, we study the quench dynamics,
\begin{equation}
    \ket{\Psi(t)}
    =
    \exp(-i\hat{H}t)
    \ket{\Psi_{\text{GZ}}}
\end{equation}
under the following Hamiltonian,
\begin{equation}
    \hat{H}=\hat{H}_0
    +\hat{H}_h
    +\hat{H}_\lambda
    +\hat{H}_{P},
    \label{Eq::Hamiltonian}
\end{equation}
Here, the first, second, and third terms, $\hat{H}_0$, $\hat{H}_h$, and $\hat{H}_\lambda$, represent the XYZ Hamiltonian, the helical magnetic field, and the non-Hermitian spin-relaxation term, respectively, as defined in Eq.\,\ref{Eq::XYZ_Hamiltonian}, Eq.\,\ref{Eq::HelicalMagneticField}, and Eq.\,\ref{Eq::non-Hermitian2}, respectively.
The scar state in Eq.\,\ref{Eq::GZScar} is stabilized by the helical field ($h>0$) and spin relaxation ($\lambda>0$).
Additionally, we incorporate a perturbative term $\hat{H}_P$. 
Specifically, we consider two types of perturbations—onsite and hopping—defined as follows,
\begin{equation}
\begin{split}
    &\hat{H}_P^{(1)}
    =\mu \sum_{n=1}^N \opS{x}{n}\,,
    \\
    &\hat{H}_P^{(2)}
    =
    \mathfrak{t} \sum_{n=1}^N 
    \left(
        \opS{+}{n} \opS{-}{n+1}+
        \opS{-}{n} \opS{+}{n+1}
    \right)\,.
\end{split}
\end{equation}
The GZ state decays under these perturbations to the Hamiltonian.

We implement the iTEBD technique to study the dynamics of such spin system using the TenPy package\,\cite{tenpy2024} for an infinite system size without any finite-size effect.
There are two parameters that control the accuracy of the iTEBD simulation. 
The first is the Trotter time step $\delta\tau$, which discretizes the continuous time evolution, and the second is the bond dimension $\chi$ of the matrix product state (MPS), which determines how accurately the wavefunction is approximated at a given time.
We employ two sets of iTEBD parameters, $(\delta\tau = 10^{-3}, \chi = 120)$ and $(\delta\tau = 5 \times 10^{-4}, \chi = 140)$, indicated by dots and solid lines, respectively, in Fig.\,\ref{fig::Dynamics}(a)–(e). 
By comparing the time-evolved observables obtained from simulations using these two parameter sets, we find that the results are well converged up to time $t = 15$ for spin $S = 1$.
For an infinite system, the observable we evaluate is the largest eigenvalue of the transfer matrix constructed from the GZ state and the time-evolved matrix product states (MPS).
This quantity can be interpreted as an effective fidelity $\mathcal{F}_{\text{GZ}}$ for an infinite system, since the conventional fidelity between two distinct quantum states vanishes identically in the thermodynamic limit.
Finally, unless stated otherwise, the following Hamiltonian parameters are set to the default values:
$\kappa = 0.5$, $\gamma = 0.8$, $q = 4K(\kappa)/5$, $S = 1$, $h=0.0$, $\mu=0.3$, and $\mathfrak{t}=0.0$.
All Hamiltonian coefficients are scaled in units of $J_y$, i.e.\, we set $J_y = 1$.

In Fig.\,\ref{fig::Dynamics}(a) and (b), the fidelity $\mathcal{F}_{\text{GZ}}$ is plotted against time in presence of onsite ($\mu=0.3$) and hopping perturbation  ($\mathfrak{t}=0.3$) respectively.
In the absence of the non-Hermitian term ($\lambda = 0$), the decay of fidelity is shown by the black curve.
In particular, in Fig.\,\ref{fig::Dynamics}(b), the fidelity for $\lambda=0$ is shown only up to $t=7$ due to a lack of convergence at later times.
The resulting dynamics closely resemble those of a generic thermal state of a chaotic system\,\cite{ThoulessTime}. 
Specifically, the fidelity initially exhibits an exponential decay, followed by a crossover to a power-law decay at later times.
This behavior also indicates that the simulation time remains shorter than the Fock-space Thouless time\,\cite{ThoulessTime}.
Moreover, since the GZ state constitutes a scar state in the spin-$1$ XYZ chain, its decay under perturbations is expected to be very slow\,\cite{SlowThermalization}.
This expectation is indeed borne out for the onsite perturbation.
In contrast, the dynamics become significantly faster in the presence of hopping perturbations. 
This distinction arises because the stability and decay of scar states depend sensitively on the nature of the perturbation.
In particular, not only the form but even the sign of a given perturbation can qualitatively influence the decay dynamics, leading to asymmetric decay behavior of such scar states\,\cite{AssyemtricDecay}.
Upon introducing non-Hermiticity ($\lambda = \pm 0.2$ and $\pm 0.4$), the power-law decay regime is replaced by a saturation regime with either finite or vanishing fidelity, as illustrated by the red and blue curves in Fig.\,\ref{fig::Dynamics}(a),(b).
In particular, positive non-Hermiticity slows down the decay rate in the exponential regime, leading to an earlier onset of a steady state with finite fidelity relative to the GZ state.
This saturation indicates the stabilization and revival of the GZ state, suggesting that in the strong non-Hermitian limit $\lambda \rightarrow +\infty$, the GZ state becomes fully stable against perturbations.
We further present phase diagrams based on the fidelity $\mathcal{F}_{\text{GZ}}$ evaluated at the saturation time $t=15$, in the parameter spaces $(\mu,\lambda)$ and $(\mathfrak{t},\lambda)$ for spin values $S=1$ and $S=3/2$, as detailed in Appendix\,\ref{Appendix::Sec::AlternativeStabilizers}.

Next, we inspect whether the application of an external helical magnetic field is able to stabilize the GZ scar state.
In Fig.\,\ref{fig::Dynamics}(c), we have plotted fidelity as a function of time for different values of magnetic field in the presence of onsite perturbation $\mu=0.3$.
The qualitative features of the dynamics remain the same in both the absence and presence of the helical magnetic field. 
The key distinction is that, for positive (negative) field strength, the exponential decay regime shortens (lengthens) in time, with the algebraic decay setting in at earlier (later) times.
Thus, a helical magnetic field of positive strength provides partial stabilization of the GZ state over a finite time window.
However, unlike the non-Hermitian case, it does not lead to a steady state with finite fidelity.

Additionally, in Fig.\,\ref{fig::Dynamics}(d), we illustrate the formation of GZ state with $q=4K(\kappa)/6$ starting from several simple product states, namely $\ket{1111\ldots}$, $\ket{1\bar{1}1\bar{1}\ldots}$, and $\ket{1\bar{1}\bar{1}1\bar{1}\bar{1}\ldots}$, under the Hamiltonian defined in Eq.\,\ref{Eq::Hamiltonian} with non-Hermiticity $\lambda=0.4$.
Here, the numbers within the kets denote the local $S^z$ quantum number at each site, while a bar over a number indicates its negative value.
In other words, here we study the following quench dynamics,
\begin{equation}
    \ket{\Psi(t)}
    =
    e^{-i\hat{H}t}
    \ket{\Psi_{\text{prod}}},
\end{equation}
where we measure the fidelity of the desired GZ state in $\ket{\Psi(t)}$ with the system initiated at the product states $\ket{\Psi_{\text{prod}}}=\ket{1111\ldots}$, $\ket{1\bar{1}1\bar{1}\ldots}$, and $\ket{1\bar{1}\bar{1}1\bar{1}\bar{1}\ldots}$.
We compare the dynamics in the absence (dense curves) and presence (transparent curves) of onsite perturbations.
Except for the fully polarized state $\ket{1111\ldots}$, which has a finite initial overlap with the GZ state ($\mathcal{F}_\text{GZ}=0.5$), the fidelity $\mathcal{F}_\text{GZ}$ starts from zero at $t=0$ for all other initial states.
In the absence of perturbations, the fidelity reaches unity (zero) for positive (negative) $\lambda$ within a finite time, signaling the complete formation (annihilation) of the GZ state.
In contrast, when an onsite perturbation($\mu=0.3$) is present, the GZ state forms only partially for positive non-Hermitian strength $\lambda$.

We next address whether an arbitrary non-eigen product state can be stabilized via such a non-Hermitian mechanism.
The non-Hermitian spin-relaxation construction in Eq.\,\ref{Eq::non-Hermitian2} is not specific to the GZ state and can, in principle, be extended to other product states.
However, for generic non-integrable and chaotic many-body Hamiltonians—such as the XYZ model with $S \geq 1$—it is highly atypical for high-energy eigenstates to exhibit a simple product-state structure, as would be required for their identification as quantum many-body scar states.
Therefore, one does not, in general, expect an arbitrary non-eigen product state to be stabilizable.
For example, for the product states $\ket{\Psi_{\text{prod}}}=\ket{1111\ldots}$, $\ket{1\bar{1}1\bar{1}\ldots}$, $\ket{1\bar{1}\bar{1}1\bar{1}\bar{1}\ldots}$, the following non-Hermitian terms are suppossed to stabilize the states, respectively,
\begin{equation}
\begin{split}
&\hat{H}_\lambda^{(1)}
=i\lambda\sum_n \hat{S}_n^z,\,
    \hat{H}_\lambda^{(1)}
    =i\lambda\sum_n (-1)^{(n+1)}\hat{S}_n^z,
    \\
    &\hat{H}_\lambda^{(3)}
    =i\lambda\sum_{n=1,4,7,\ldots} \hat{S}_n^z
    -i\lambda\sum_{n=2,3,5,6,\ldots} \hat{S}_n^z.
    \label{Eq::NonHermitianNonEigenState}
\end{split}
\end{equation}
To examine the impact of these non-Hermitian terms on the corresponding product states, we replace the $H_\lambda$ term in Eq.\,\ref{Eq::Hamiltonian} with these non-Hermitian terms and analyze the resulting dynamics within the following quench protocol,
\begin{equation}
    \ket{\Psi(t)}
    =
    e^{-i\hat{H}t}
    \ket{\Psi_{\text{prod}}},
\end{equation}
where we evaluate the fidelity $\mathcal{F}_{\text{prod}}$ of the time-evolved state relative to the initial product states $\ket{\Psi_{\text{prod}}}$.
However, since not all product states are eigenstates of the interacting XYZ many-body Hamiltonian, such stabilization is not generally expected to succeed.
The results, shown in Fig.\,\ref{fig::Dynamics}(f), confirm this expectation: except for the ferromagnetic state, the other product states cannot be stabilized.
Interestingly, the ferromagnetic state $\ket{1111\ldots}$ can still be stabilized for $\lambda > 0$ (see the solid blue curve), which can be attributed to its large overlap with the GZ scar subspace. This is consistent with the observation in Fig.\,\ref{fig::Dynamics}(e), where the overlap between the ferromagnetic state and the GZ state is already significant ($\mathcal{F}_{\text{GZ}} \approx 0.5$) at $t = 0$.
This suggests that the ferromagnetic state can be interpreted as a scar state of the XYZ system.
However, while the GZ state is an exact scar state of the XYZ system, the ferromagnetic state, although also a scar state, is not an exact one.
Nevertheless, this indicates that the ferromagnetic state, although a non-eigen product state, can also be stabilized via non-Hermitian spin relaxation due to its finite overlap with the GZ scar subspace.
A systematic analysis of the overlap between the GZ scar subspace and the product states $\ket{\Psi_{\text{prod}}}=\ket{1111\ldots}$, $\ket{1\bar{1}1\bar{1}\ldots}$, $\ket{1\bar{1}\bar{1}1\bar{1}\bar{1}\ldots}$ is presented in Appendix.\,\ref{Appendix::Sec::AlternativeStabilizers}.

Finally, we show that the stabilization of the GZ state persists even for finite system sizes and under local perturbations.
Specifically, we consider a nine-site non-Hermitian spin-$1$ XYZ chain with periodic boundary conditions and study the dynamics of a system initialized in the GZ state under either of the following local perturbative terms,
\begin{equation}
    \hat{H}_{P}^{(3)}=\mu_1 \opS{z}{1},\,\,
    \hat{H}_{P}^{(4)}=\mu_2 \opS{+}{1} \opS{-}{1} \opS{-}{2} \opS{+}{2}.
\end{equation}
In Fig.\,\ref{fig::Dynamics}(f), we observe that for positive non-Hermiticity $\lambda = 0.4$ (see black curves), the system reaches a steady state with finite fidelity relative to the GZ state, thereby partially stabilizing it and demonstrating the validity of the stabilization scheme for finite-size systems in the presence of local perturbations.
The quantum dynamics of finite-size closed quantum systems are simulated using the QuTiP package\,\cite{QuTip}.

\section{Physical Realization}
\label{Sec::Physical Realization}

\begin{figure}[t]
\centering
\includegraphics[width=0.45\textwidth]{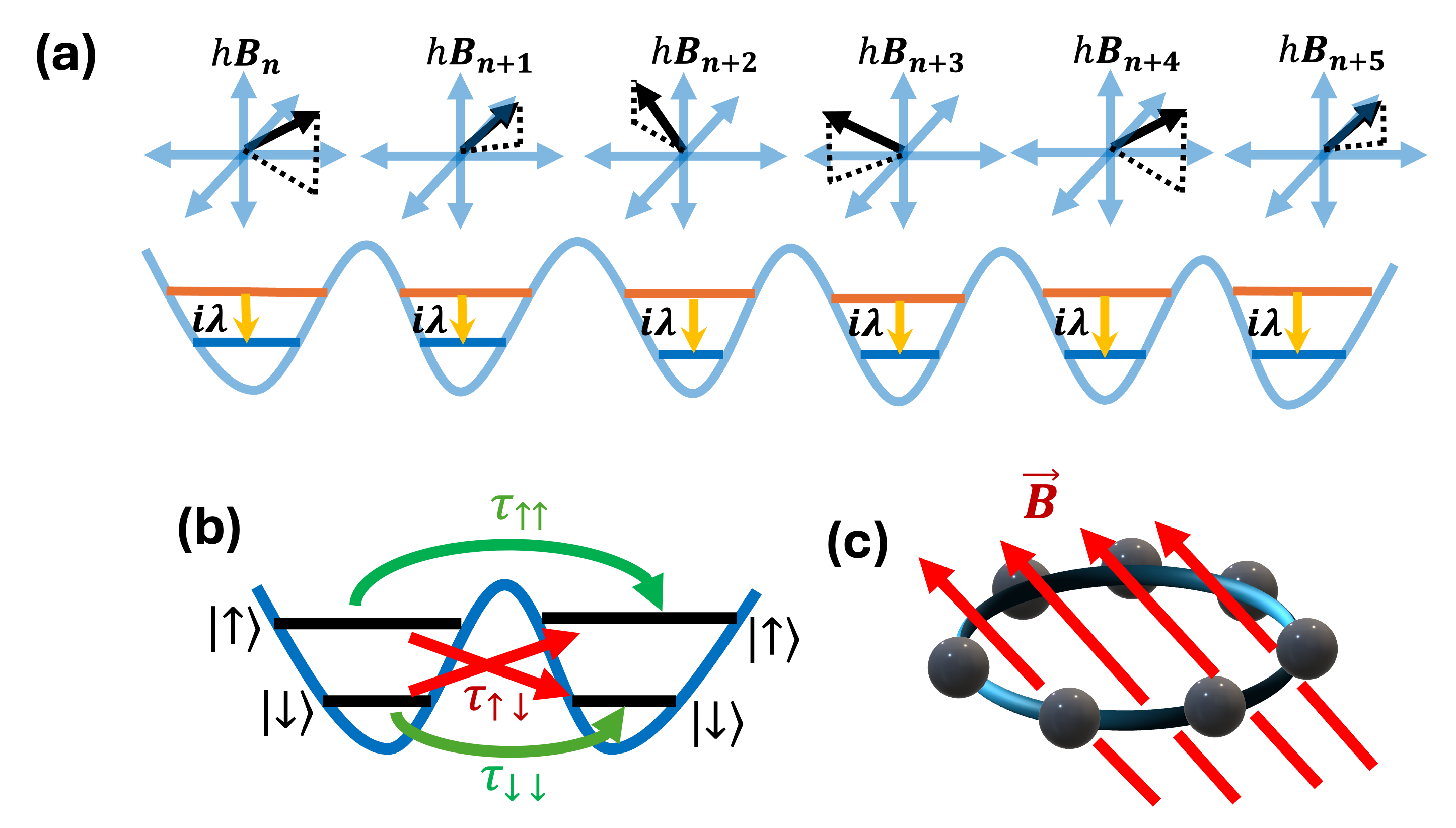}
\caption{
(a) Schematic of the experimental arrangement using an optical lattice.
A periodic potential is illustrated, within which the atoms are confined.
A site-dependent magnetic field $h\boldsymbol{B}_n$ is applied at each site $n$ along the $S^{z'}_n$ direction, aligned with the local magnetization of the GZ state.
This field lifts the spin degeneracy, yielding lower-energy down-spinon (orange) and higher-energy up-spinon (blue) states.
Coupling to the environment induces relaxation from the down-spinon to the up-spinon state at a rate $\lambda$.
(b) Schematic of the hopping coefficients in the Hubbard model. 
In the absence of a magnetic field, the spin-up and spin-down degrees of freedom are degenerate; they are depicted as energetically separated here for illustrative purposes only.
(c) Schematic illustration of a ring-lattice setup in which atoms (shown as black spheres) realize a non-Hermitian XXZ model under a uniform magnetic field tilted at a finite angle relative to the plane of the ring. 
In this configuration, the uniform magnetic field effectively appears as a helical field in the local reference frame of each lattice site. In such a setup, a GZ state with wavevector $q=2\pi/N$ can be stabilized for any arbitrary value of $\gamma$.
}
\label{fig::Experiment}
\end{figure}

\subsection{Spin Relaxation}
To realize such spin relaxation, a helical magnetic field is applied, which lifts the degeneracy of the local spinon levels and splits them into up- and down-spinon states.
The GZ state, being the vacuum state of the up-spinon state, can be stabilized by the relaxation process that corresponds to the decay from the down-spinon state to the up-spinon state, with the coefficient $\lambda$ characterizing the decay rate (see Fig.\,\ref{fig::GZ_Schematic}(b)).
Such spin relaxation may arise through several mechanisms depending on the environmental bath, including spontaneous decay, Purcell-enhanced decay, and spin–lattice relaxation.
Spontaneous spin relaxation in a vacuum is typically negligible due to its extremely long lifetime. 
Resonator engineering can significantly enhance the relaxation rate via the Purcell effect by modifying the electromagnetic density of states near the energy corresponding to the splitting between the up- and down-spinon states, determined by the magnitude of the helical magnetic field\,\cite{PurcellEffect1, PurcellEffect2, PurcellEffect3}.
Such modifications of the electromagnetic density of states can be achieved by realizing the spin system inside an electromagnetic resonator.
Notably, the implementation of optical lattices within electromagnetic resonators has been demonstrated in experiments\,\cite{RingOpticalLattice}.
Alternatively, an onsite magnetostriction can give rise to such a spin-relaxation process.


In particular, we consider a system with phonon-mediated spin relaxation to explicitly relate the non-Hermitian coefficient $\lambda$ as the spin-relaxation decay rate. 
We consider a system with dominant onsite magnetostriction, a general form of which is given as follows,
\begin{equation}
    \hat{H}_{mc}
    =
    \sum_{n=1}^N \sum_{pqr} \gamma_{pqr} \hat{u}_n^p \opS{q}{n} \opS{r}{n}\,,
    \label{Eq::Magnetostriction}
\end{equation}
where, indices $(p,q,r)\in\left\lbrace x,y,z\right\rbrace$ denoting the Cartesian components.
We further assume that the one-dimensional spin chain is embedded in a substrate or a larger lattice environment, such that the emitted phonons dissipate into the surrounding medium and do not re-excite the spins of the system.
Here, $\gamma_{pqr}$ represents the magnetostriction coefficient, while $\hat{u}_n^p$ denotes the lattice displacement operator of the $n$-th site.
The position operator in terms of phononic normal modes can be expressed as,
\begin{equation}
    \hat{u}_n^p 
    =
    \frac{1}{\sqrt{N}}
    \sum_{km} 
    \left[e_{km}\right]_p
    \sqrt{\frac{\hbar}{2M\omega_{km}}}
    \left(
    \hat{\gamma}_{km}^\dagger 
    +\hat{\gamma}_{-km}
    \right)
    e^{inka}\,,
    \label{Eq::PhononModes}
\end{equation}
where $e_{km}$, $\omega_{km}$, and $\hat{\gamma}_{km}$ are the phonon eigenvector, eigenvalue, and annihilation operator of the $m$-th phonon mode at $k$-th lattice momentum, respectively.
$M$, $a$ and $[e_{km}]_p$ are the atomic mass, lattice constant and $p$-th element of $e_{km}$ vector, respectively.
According to Fermi's golden rule, the spin relaxation rate,
\begin{equation}
\begin{split}
    |\boldsymbol{\lambda}_n|=&\frac{2\pi}{\hbar}
    \sum_{\substack{kmn_\uparrow n_\downarrow \\ n_\uparrow +n_\downarrow=2S}} 
    \left|
    \bra{km}\bra{n_\uparrow+1, n_\downarrow-1}
    \left[\hat{H}_{mc}\right]_n
    \ket{n_\uparrow, n_\downarrow}
    \ket{0}
    \right|^2
    \\
    &\qquad\qquad\qquad\qquad\qquad\qquad
    \times\frac{\delta(\omega_{km}-h)}
    {2S},
\end{split}
\label{Eq::FermisGoldenRule}
\end{equation}
where the states $\ket{n_\uparrow,n_\downarrow}$ and $\ket{k m}$ correspond to configurations with $n_\uparrow$ ($n_\downarrow$) up-spin (down-spin) spinons and a phonon eigenmode $(k,m)$, respectively. 
The factor $2S$, in the denominator, is introduced to average the decay rates over all spinon configurations.
Using Eq.\,\ref{Eq::Magnetostriction}, Eq.\,\ref{Eq::PhononModes} and Eq.\,\ref{Eq::FermisGoldenRule}, we obtain,
\begin{equation}
\begin{split}
&|\boldsymbol{\lambda}_n|=\frac{\pi}{2NMS}
\sum_{\substack{kmn_\uparrow n_\downarrow\\pqr}}
c_{pqr} \left|
\bra{n_\uparrow+1, n_\downarrow-1}
\opS{\beta}{n} \opS{\gamma}{n}
\ket{n_\uparrow, n_\downarrow}\right|^2
\\
&\qquad\qquad\qquad\qquad\qquad
\times \delta(\omega_{km}-h)\,\delta(n_\uparrow+n_\downarrow-2S)\,,
\end{split}
\label{Eq::FermisGoldenRule2}
\end{equation}
where $c_{pqr}=\left|\gamma_{pqr}\left[e_{km}\right]_p\right|^2$.
We emphasize that the spinon states appearing in this expression are defined in the locally rotated (primed) basis in which $z^\prime$-axis is defined along the local spin direction in Fig.\,\ref{fig::GZ_Schematic}, while the spin operators are written in the laboratory frame. As a result, the coefficient $|\boldsymbol{\lambda}_n|$ of the non-Hermitian term acquires an explicit site dependence.
The expression Eq.\,\ref{Eq::FermisGoldenRule2} suggests that the spin-relaxation coefficient $\lambda$ can be tuned via the spin–lattice coupling coefficient $\gamma_{pqr}$, the energy gap $h$ between the up- and down-spinon states (which can be controlled by an external magnetic field), and the phonon density at the corresponding energy.
Moreover, in optical lattice setups, the phonon density of states can be engineered and controlled\,\cite{PhononOpticalLattice1,PhononOpticalLattice2}.

\subsection{Hubbard Model}
In both solid-state and cold-atom systems, spin models are typically realized as effective low-energy descriptions of underlying Hubbard models\,\cite{XXZ_chain1, XXZ_chain2, ColdAtomExperiment}. 
In this section, we identify the underlying Hubbard model required to realize a non-Hermitian spin-$1$ XYZ model with a helical magnetic field. 
We consider the following two-orbital fermionic Hubbard model at half-filling,

\begin{widetext}

\begin{equation}
\begin{split}
    \hat{H}&=
    -\sum_{j=1}^N \sum_{m=1,2} \left(
    \tau_{\uparrow} \hat{c}^\dagger_{j,m,\uparrow} \hat{c}_{j+1,m,\uparrow} 
    +\tau_{\downarrow} \hat{c}^\dagger_{j,m,\downarrow} \hat{c}_{j+1,m,\downarrow}
    +\tau_{\uparrow\downarrow} \hat{c}^\dagger_{j,m,\uparrow} \hat{c}_{j+1,m,\downarrow}
    +\text{h.c.}
    \right)
    +U\sum_{j=1}^N \sum_{m=1,2}\hat{n}_{j,m,\uparrow} \hat{n}_{j,m,\downarrow}
    \\
    &+U^\prime \sum_{j=1}^N \sum_{\sigma,\sigma^\prime}\hat{n}_{j,1,\sigma} \hat{n}_{j,2,\sigma^\prime}
    -J_H\sum_j \hat{\boldsymbol{s}}_{j,1}\cdot \hat{\boldsymbol{s}}_{j,2}
    -\frac{1}{2} \sum_{j=1}^N \sum_{m} \hat{c}^\dagger_{j,m} \boldsymbol{h}_j\cdot\boldsymbol{\sigma} \hat{c}_{j,m}
    +\frac{i}{2} \sum_{j=1}^N \sum_{m} \hat{c}^\dagger_{j,m} \boldsymbol{\lambda}_j\cdot\boldsymbol{\sigma} \hat{c}_{j,m},
\end{split}
\label{Eq::HubbardModel}
\end{equation}

\end{widetext}
where $\hat{c}_{j,m,\sigma}$
denotes fermionic annihilation operator for site $j$, orbital $m$ and spin $\sigma$.
Moreover, the coefficients $(\tau_{\uparrow},\,\tau_{\downarrow},\,\tau_{\uparrow\downarrow})$, $U$, $U^\prime$, and $J_H$ represent spin-dependent hopping amplitudes, onsite intra-orbital Hubbard repulsion, onsite inter-orbital Hubbard repulsion, and inter-orbital Hund's coupling, respectively.
The operator $\hat{c}_{j,m}=(\hat{c}_{j,m,\uparrow},\hat{c}_{j,m,\downarrow})^T$.
Additionally, the vectors $\boldsymbol{h}_j$ and $\boldsymbol{\lambda}_j$ are parallel to each other; the former represents an external magnetic field, while the latter denotes the non-Hermitian spin-relaxation.
We also assume that the magnitudes $|\boldsymbol{h}_j|=h$ and $|\boldsymbol{\lambda}_j|=\lambda$ are site independent.
The Hund's coupling $J_H$ splits the triplet and singlet sectors, making the triplets the low-lying energy states at site-$j$,
\begin{equation}
\begin{split}
\ket{t_1}_j=\ket{\uparrow_1\uparrow_2}_j,\,
\ket{t_{\bar{1}}}_j=\ket{\downarrow_1 \downarrow_2}_j,
\\
\ket{t_0}_j=\frac{1}{\sqrt{2}}
\left(\ket{\uparrow_1\downarrow_2}_j+\ket{\downarrow_1\uparrow_2}_j\right),\,
\end{split}
\end{equation}
where, $\uparrow_m$ (or $\downarrow_m$) denotes a spin-$1/2$ degrees of freedom associated with the orbital $m$.
The states $\ket{t_1}$, $\ket{t_0}$, and $\ket{t_{\bar{1}}}$ represent the spin-triplet states with $S^z$ quantum number $+1$, $0$, and $-1$, respectively.
The triplet states thus form an effective spin-$1$ degree of freedom.
In the presence of the magnetic field this triplets further split into the following states,
\begin{equation}
\begin{split}
    &\ket{-h}_j
    =
    e^{-i\delta_n} \cos^2\left(\frac{\chi_j}{2}\right) \ket{t_1}_j
    + \frac{1}{\sqrt{2}} \sin\left(\chi_j\right) \ket{t_0}_j
    \\
    &\qquad\qquad\qquad\qquad
    + e^{i\delta_j} \sin^2\left(\frac{\chi_j}{2}\right) \ket{t_{\bar{1}}}_j,
    \\
    &\ket{0}_j=-\frac{e^{-i\delta_j}}{\sqrt{2}} \sin(\chi_j) \ket{t_1}_j
    +\cos(\chi_j) \ket{t_0}_j
    \\
    &\qquad\qquad\qquad\qquad
    +\frac{e^{i\delta_j}}{\sqrt{2}} \sin(\chi_j) \ket{t_{\bar{1}}}_j,
    \\
    &\ket{+h}_j=e^{-i\delta_j} \sin^2\left(\frac{\chi_j}{2}\right) \ket{t_1}_j
    - \frac{1}{\sqrt{2}} \sin\left(\chi_j\right) \ket{t_0}_j
    \\
    &\qquad\qquad\qquad\qquad
    + e^{i\delta_j} \cos^2\left(\frac{\chi_j}{2}\right) \ket{t_{\bar{1}}}_j,
\end{split}
\end{equation}
where $\chi_j$ and $\delta_j$ represent the polar (or zenith) and azimuthal angles, respectively, specifying the direction of the magnetic field: $h^x_j/h=\sin\chi_j\cos\delta_j$, $h^y_j/h=\sin\chi_j\sin\delta_j$, $h^z_j/h=\cos\chi_j$.  
The energies of the states $\ket{-h}_j$, $\ket{0}_j$, and $\ket{+h}_j$ vary linearly with the external magnetic field as $-h$, $0$, and $h$, respectively.
These states can be interpreted in terms of spinons as follows: $\ket{-h}_j$ corresponds to a two–down-spinon state, $\ket{0}_j$ to a one–up–one–down spinon state, and $\ket{+h}_j$ to a two–up-spinon state.
The non-Hermitian term arises basically due to spin relaxation from the down spinon states to the up spinon states, as depicted schematically in Fig.\,\ref{fig::GZ_Schematic}(b).
The schematic of the Hubbard model is depicted in Fig.\,\ref{fig::Experiment}(a) and (b).
Finally, at half filling and in the strong-coupling limit $\tau_{\sigma\sigma^\prime}/U$, $J_H/U$, $h/U\ll 1$, the Hubbard model perturbatively maps to spin-$1$ non-Hermitian XYZ model with helical magnetic field via a Schrieffer–Wolff transformation and projection to triplet states for $h<J_H$ (ignoring constant terms),
\begin{equation}
\begin{split}
    \hat{H}=\sum_{n=1}^N &\left[
    J \left(\opS{1}{n}\opS{1}{n+1}
    + \opS{2}{n}\opS{2}{n+1}
    \right)
    +\Delta \opS{3}{n}\opS{3}{n+1}
    \right.
    \\
    &\qquad\left.
    +J_{13} \opS{1}{n} \opS{3}{n+1}
    +J_{13} \opS{3}{n} \opS{1}{n+1}
    \right]
    \\
    &
    -\sum_{n=1}^N \boldsymbol{h}_n\cdot\hat{\boldsymbol{S}}_n
    +i\sum_{n=1}^N \boldsymbol{\lambda}_n\cdot\hat{\boldsymbol{S}}_n\,
    \\
    &+
    \mathcal{O}\left(\frac{\tau^2 h}{U}\right)
    +\mathcal{O}\left(\frac{\tau^2 J_H}{U}\right)
    +\mathcal{O}\left(\frac{\tau_{\uparrow\downarrow}^2}{U}\right)
    ,
\end{split}
\end{equation}
where,
\begin{equation*}
    \begin{split}
        &J=(2/U)\tau_\uparrow \tau_\downarrow\,,
        \Delta=(1/U) (\tau_\uparrow^2+\tau_\downarrow^2)\,,
        \\
        &
        J_{13}=(2/U)\tau_{\uparrow\downarrow}(\tau_\uparrow-\tau_\downarrow).
    \end{split}
\end{equation*}
We note that, for $J_{13}$ to be significant relative to the perturbative contribution $\mathcal{O}(\tau_{\uparrow\downarrow}^2/U)$, the condition $|\tau_\uparrow - \tau_\downarrow| \gg \tau_{\uparrow\downarrow}$ must be satisfied.
Next, using the following $SO(3)$ rotation,
\begin{equation}
\begin{split}
    \begin{pmatrix}
        \opS{1}{n} \\ \opS{2}{n} \\ \opS{3}{n}
    \end{pmatrix}
    &=
    \begin{pmatrix}
        -\sin\theta & \cos\theta & 0 \\
        0 & 0 & 1 \\
        \cos\theta & \sin\theta & 0
    \end{pmatrix}
    \begin{pmatrix}
        \opS{x}{n} \\ \opS{y}{n} \\ \opS{z}{n}
    \end{pmatrix}\,,
    \\
    &\text{with  } \tan(2\theta) = \frac{2J_{13}}{J-\Delta}\,,
\end{split}
\end{equation}
we obtain the spin-exchange Hamiltonian, as,
\begin{equation}
\begin{split}
    \hat{H}=\sum_{n=1}^N \left[
    J_x \opS{x}{n}\opS{x}{n+1}
    + J_y \opS{y}{n}\opS{y}{n+1}
    + J_z \opS{z}{n} \opS{z}{n+1}
    \right]
    \\
    -h\sum_{n=1}^N \boldsymbol{B}_n \cdot \boldsymbol{\hat{S}}_n
    +i\lambda \sum_{n=1}^N \boldsymbol{B}_n \cdot \boldsymbol{\hat{S}}_n \,,
\end{split}
\label{Eq::DerivedNonHermitian}
\end{equation}
where,
\begin{equation}
\begin{split}
    &J_x=J\,,
    \\
    &J_y=\frac{1}{2} \left[J+\Delta+\sqrt{(J-\Delta)^2+4J_{13}^2}\right]\,,
    \\
    &J_z=\frac{1}{2} \left[J+\Delta-\sqrt{(J-\Delta)^2+4J_{13}^2}\right]\,,
    \\
    & B_n^x=\frac{1}{h} (-h_n^x\sin\theta+h_n^z\cos\theta).
    \\
    &B_n^y=\frac{1}{h} (h_n^x \cos{\theta}+h_n^z\sin\theta)\,, 
    B_n^z=\frac{h_n^y}{h}\,,
    \\
    & h=|\boldsymbol{h}_n|\,,\lambda=|\boldsymbol{\lambda}_n|
\end{split}
\end{equation}
Therefore, we demonstrate the connection between the non-Hermitian spin-exchange Hamiltonian for spin-$1$ in Eq.\,\ref{Eq::DerivedNonHermitian} with the non-Hermitian Hubbard model Eq.\,\ref{Eq::HubbardModel}.
For higher spin values, the non-Hermitian XYZ model can similarly be obtained from a multiorbital fermionic Hubbard model with Hund's coupling.

\subsection{Helical Magnetic Field}

One of the main experimental challenges is realizing a helical magnetic field at each lattice site.
This difficulty can be circumvented, for example, in the XXZ model by implementing the system on a ring geometry and applying a uniform magnetic field tilted with respect to the plane of the ring (see schematic in Fig.\,\ref{fig::Experiment}(c)). 
In this setup, the uniform magnetic field effectively appears as a helical field in the local reference frame of each lattice site.
At the same time, this resolves issues related to the thermodynamic limit, since periodic boundary conditions allow the GZ state to be realized as an eigenstate.
Such ring geometries have already been realized experimentally\,\cite{RingOpticalLattice,RingLattice1,RingLattice2}. 
In this configuration, the GZ state with wavevector $q=2\pi/N$ can be stabilized for arbitrary values of $\gamma$.


\section{Equivalent Lindbladian Description}

\begin{figure}[t]
\centering
\includegraphics[width=0.35\textwidth]{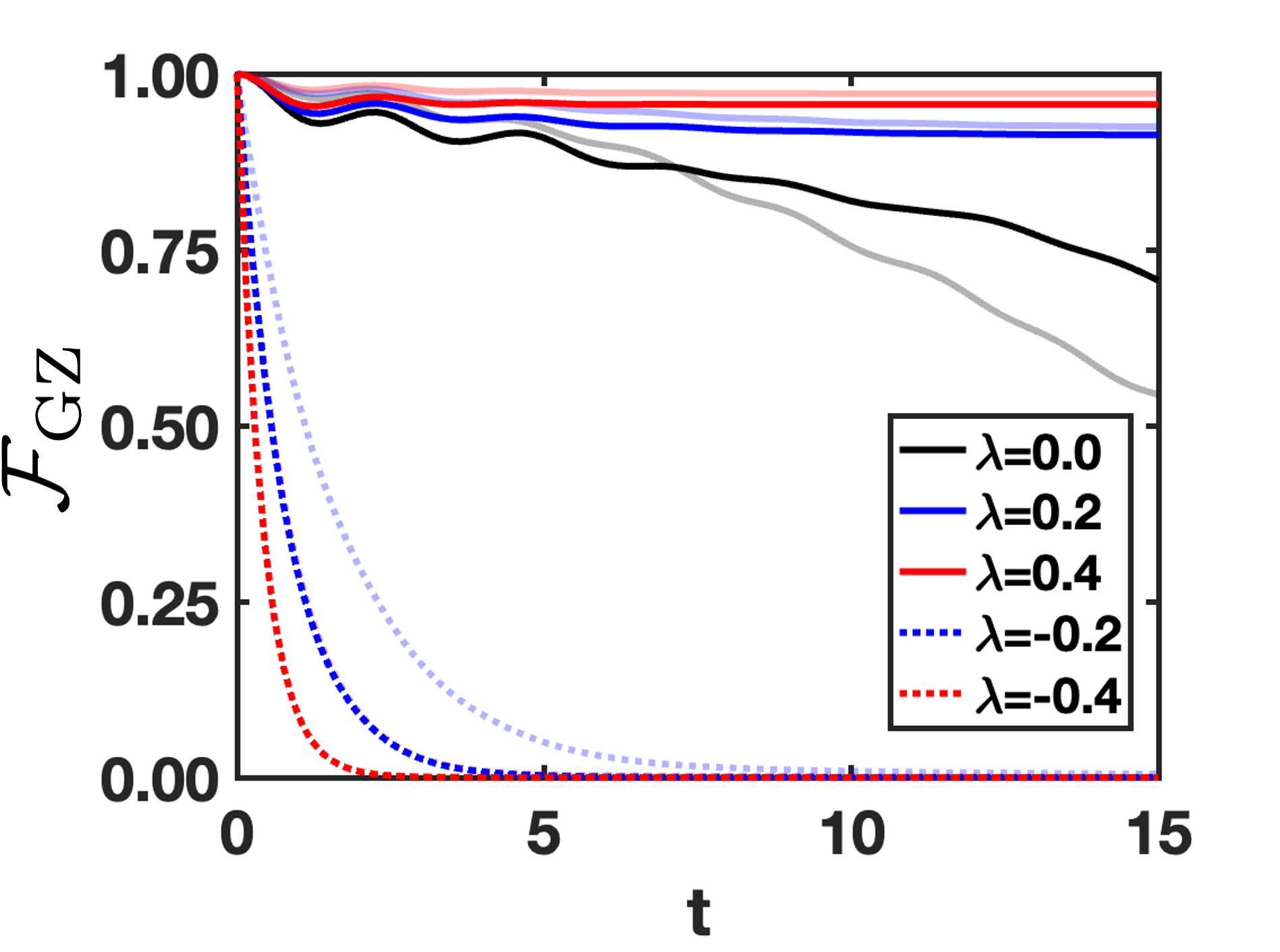}
\caption{
Lindbladian dynamics governed by the system Hamiltonian $\hat{H}_0$, in the presence of an onsite perturbation $\mu = 0.3$, and with the Lindblad jump operator defined in Eq.\,\ref{Eq::JumpOperator}.
Results are shown for finite system sizes $N = 3$ (transparent lines) and $N = 6$ (dense lines).
The parameter $q$ is fixed to $q = 4K(\kappa)/3$, while the remaining system parameters are set to $\kappa = 0.5$, $\gamma = 0.8$, and $S = 1$.
The simulation result for $\lambda = -0.4$ and system size $N = 3$ is not visible, as it lies below the blue dashed curve corresponding to $\lambda = -0.2$ for the system size $N = 6$.
}
\label{fig::Lindbladian}
\end{figure}

A non-Hermitian Hamiltonian provides an effective description of an open quantum system, and so does the Lindbladian formalism. 
Each approach has its own advantages and limitations. 
For instance, a non-Hermitian Hamiltonian is well suited to capture the short-time dynamics of an open system, whereas the Lindbladian approach can accurately describe the long-time dynamics, provided that the system–environment coupling remains weak.
Interestingly, every Lindbladian dynamics has a corresponding effective non-Hermitian description, which remains valid for short times until the effects of quantum jump operators become dominant\,\cite{LindbladToNH1, LindbladToNH2, LindbladToNH3}.
For example, the Lindblad master equation has the following form,
\begin{equation}
    \dot{\rho}
    =
    -i\left[\hat{H}_0, \rho \right]
    +\sum_j \Gamma_j 
    \left(
        L_j \rho L_j^\dagger
        -\frac{1}{2}
        \left\lbrace
            L_j^\dagger L_j, \rho
        \right\rbrace
    \right),
    \label{Eq::LindbladMasterEquation}
\end{equation}
where $\left\lbrace L_j \right\rbrace$ denotes the set of quantum jump operators, and the associated gain/decay rates $\Gamma_j$ must be positive in order for the Lindbladian dynamics to define a completely positive map.
By neglecting the quantum jump operator of the form $L_j \rho L_j^\dagger$, the Lindblad master equation Eq.\,\ref{Eq::LindbladMasterEquation} reduces to the von-Neumann equation of motion with an effective non-Hermitian Hamiltonian,
\begin{equation}
    \hat{H}_{\text{eff}}=\hat{H}_0
    -\frac{i}{2}\sum_j \Gamma_j \hat{L}_j^\dagger \hat{L}_j.
\end{equation}
However, the converse is not true: not every non-Hermitian Hamiltonian admits a corresponding Lindblad master equation.
A Lindbladian description of a non-Hermitian Hamiltonian exists only when the anti-Hermitian part of the non-Hermitian Hamiltonian is negative semidefinite, ensuring that the resulting Lindbladian evolution defines a completely positive map ($\Gamma_j \geq0$).
In the present case, the operator $\lambda \opS{z^\prime}{n}$ is neither positive nor negative semidefinite.
Consequently, the non-Hermitian terms in Eq.\,\ref{Eq::non-Hermitian1} and Eq.\,\ref{Eq::non-Hermitian2} do not admit an exact Lindbladian representation.
Nevertheless, any Hermitian operator such as $\opS{z^\prime}{n}$ can always be decomposed into positive and negative semidefinite components using eigenvalue decomposition.
We can write $\opS{z^\prime}{n}$ as,
\begin{equation}
    \opS{z^\prime}{n}=\frac{1}{2}\left[\opS{+^\prime}{n}, \opS{-^\prime}{n}\right]
    =\frac{1}{2} \left(
    \opS{+^\prime}{n} \opS{-^\prime}{n}
    -
    \opS{-^\prime}{n} \opS{+^\prime}{n}
    \right),
\end{equation}
where the first term is positive semidefinite and the second term is negative semidefinite.
By neglecting the positive semidefinite part and taking only the negative semidefinite part into account, we obtain a Lindblad master equation with,

\begin{equation}
\begin{split}
    &\Gamma_n=|\lambda|,\\
    &\hat{L}_n=\opS{+^\prime}{n},\,\, 
    \text{for }\lambda>0,\\
    &\hat{L}_n=\opS{-^\prime}{n},\,\, 
    \text{for }\lambda\leq 0.
    \label{Eq::JumpOperator}
\end{split}
\end{equation}
Next, it is important to obtain the operator $\opS{+^\prime}{n}$ in the lab frame, using the following $SO(3)$ rotation,

\begin{widetext}

\begin{equation}
    \begin{pmatrix}
            \opS{x^\prime}{n} \\
            \opS{y^\prime}{n} \\
            \opS{z^\prime}{n}
    \end{pmatrix}
    =
    \begin{pmatrix}
        -\frac{1}{N_n} \beta \sn\,(nq,\kappa)
        & \frac{1}{N_n} \alpha \cn\,(nq,\kappa)
        & 0
        \\
        -\frac{1}{N_n} \alpha \gamma \cn\,(nq,\kappa) \dn\,(nq,\kappa)
        & -\frac{1}{N_n} \beta \gamma \dn\,(nq,\kappa) \sn\,(nq,\kappa) 
        & N_n
        \\
        \alpha \cn\,(nq,\kappa) 
        & \beta \sn\,(nq,\kappa)
        & \gamma \dn\,(nq,\kappa)
    \end{pmatrix}
    \begin{pmatrix}
    \opS{x}{n}\\
    \opS{y}{n}\\
    \opS{z}{n}
\end{pmatrix},
\end{equation}

where, $N_n=\sqrt{1-\gamma^2\dn^2(nq,\kappa)}$.
Thus, the spin ladder operator becomes,

\begin{equation}
\begin{split}
    \opS{+^\prime}{n}
    =
    -\frac{1}{N_n}
    \left(
    \beta\sn\,(nq,\kappa)+i\alpha\gamma\dn\,(nq,\kappa)\cn\,(nq,\kappa)
    \right)
    \opS{x}{n}
    +\frac{1}{N_n}
    \left(
        \alpha \cn\,(nq,\kappa)
        -i\beta\gamma\dn\,(nq,\kappa)\sn\,(nq,\kappa)
        \right)
        \opS{y}{n}
        +iN_n\opS{z}{n}.
\end{split}
\end{equation}

The jump operator derived here is equivalent to the one reported by Vladislav et al., up to a global phase factor\,\cite{VladislavPopkov1, VladislavPopkov2}.

\end{widetext}

The Lindblad formalism is qualitatively equivalent to the non-Hermitian description, as both frameworks capture the same underlying physical processes.
The Lindblad approach describes a spin system weakly coupled to an environmental bath, such as a phononic or photonic reservoir. 
The corresponding jump operators induce on-site spin transitions from the down-spinon to the up-spinon state for $\lambda > 0$, and from the up-spinon to the down-spinon state for $\lambda < 0$, accompanied by the emission of a phonon or photon that carries information about the initial spin state.
Since the number of degrees of freedom of a single spin is negligible compared to that of the bath, this information is effectively irretrievable, thereby rendering the dynamics Markovian.

In Fig.\,\ref{fig::Lindbladian}, we show that the dynamics governed by a Lindbladian—partially derived from the non-Hermitian Hamiltonian—exhibit qualitatively similar behavior.
We simulate the Lindblad dynamics using the QuTiP package\,\cite{QuTip} for periodic finite-size systems ($N=3$ and $N=6$) with parameters $q=4K(\kappa)/3$, $S=1$, and $\mu=0.3$.
The fidelity between two density matrices at initial time $\rho_{\text{GZ}}$ for GZ state and at finite time $\rho(t)$ is evaluated as $\mathcal{F}_\text{GZ}=\left(\text{Tr}\sqrt{\rho(t)}\rho_{\text{GZ}}\sqrt{\rho(t)}\right)^2$.
Apart from finite-size effects, the Lindbladian dynamics in Fig.\,\ref{fig::Lindbladian} exhibit behavior very similar to the non-Hermitian dynamics shown and discussed earlier in Fig.\,\ref{fig::Lindbladian}.
Increasing the decay (or gain) rate $\Gamma_n=\lambda>0$ leads to enhanced stabilization of the GZ state. 
These results demonstrate that the Lindbladian description, derived from the non-Hermitian stabilization term, qualitatively captures the same stabilizing physics of the scar states.

\section{Conclusion}

In this work, we have investigated the stability and dynamics of Granovskii–Zhedanov (GZ) scar states in the spin-$S$ XYZ model in the presence of realistic perturbations. 
While GZ states are exact nonthermal eigenstates of the ideal Hamiltonian, we have shown that generic perturbations inevitably lead to their decay and eventual thermalization, underscoring the need for active stabilization mechanisms.
We have demonstrated that a helical magnetic field aligned with the local spin texture of the GZ state significantly enhances its lifetime by suppressing thermalization. 
However, such unitary control alone is insufficient to achieve long-time stabilization. 
To overcome this limitation, we introduced a physically motivated non-Hermitian spin-relaxation mechanism, which drives the system toward a nonequilibrium steady state with finite fidelity to the GZ state.
This mechanism provides a robust route to stabilizing GZ scars even in the presence of both onsite and hopping perturbations.

Using a combination of iTEBD simulations and exact time evolution, we systematically analyzed the dynamical behavior across both infinite and finite system sizes.
Our results show that non-Hermiticity replaces the long-time decay of fidelity with a saturation regime, indicating the emergence of a stabilized steady state. 
Furthermore, we established that while the stabilization mechanism can dynamically generate GZ states from certain initial product states, it does not generically stabilize arbitrary non-eigen product states, reflecting the special role of scar subspaces in nonintegrable systems.
We also provided a concrete physical interpretation of the non-Hermitian term in terms of spin relaxation processes arising from coupling to environmental baths, such as Purcell-enhanced spontaneous emission and spin–lattice interactions.
In addition, we constructed an equivalent Lindblad description and showed that it captures qualitatively similar stabilization dynamics, thereby connecting the non-Hermitian framework to open quantum system approaches.

Finally, we proposed experimentally relevant platforms for realizing these effects, including cold-atom systems in optical lattices and solid-state implementations described by multiorbital Hubbard models. 
Our analysis suggests that engineered dissipation, combined with tailored magnetic fields, offers a viable pathway for stabilizing many-body scar states in realistic settings.
Several open questions remain for future investigation. 
In particular, it would be interesting to explore whether spatially uniform or boundary-induced non-Hermitian terms can achieve similar stabilization, and whether a universal dissipative mechanism can stabilize an entire family of GZ states. 
Our work on non-Hermitian spin relaxation of GZ scar can be extended to the product scar states in the other spin systems.
For instance, Ref.\,\cite{FrustratedSpin2} demonstrates the emergence of singlet coverings as many-body scar states in the frustrated spin systems.
In such cases, the corresponding non-Hermitian singlet relaxation term takes the following form,
\begin{equation}
    2i\lambda\sum_{m,n\in b}
     \hat{S}_m \cdot \hat{S}_n
    =
    i\frac{3\lambda}{2} \sum_b\hat{s}_b^\dagger\hat{s}_b
    -
    i\frac{\lambda}{2} 
    \sum_{b,\alpha}
    \hat{t}^{\alpha\dagger}_b\hat{t}^\alpha_b,
    \label{Eq::SingletRelaxation}
\end{equation}
where $b$ denotes a singlet dimer bond, and $m$ and $n$ label the spin sites forming the dimer. 
The operators $\hat{s}_b^\dagger$ and $\hat{t}_b^{\alpha\dagger}$ represent the singlet and triplet creation operators, respectively, which are related to the spin operators via a bond-operator transformation\,\cite{BondOperator1, BondOperator2}.
Physically, the non-Hermitian term Eq.\,\ref{Eq::SingletRelaxation} corresponds to relaxation from the triplet manifold to the singlet state, which can be engineered by tuning the density of states of environmental photons or phonons to match the energy scale set by the Heisenberg exchange interaction of the dimer bond.
In conclusion, our results underscore the potential of non-Hermitian and dissipative engineering as powerful frameworks for the control and protection of nonthermal states in quantum many-body systems, opening new avenues for the theoretical and experimental study of quantum scars.

{\it Acknowledgments}.--- We extend thanks to Dr. Shilpi Roy for useful discussions. We acknowledge the use of the computational resources at the National Supercomputing Centre (NSCC), Singapore. 



\onecolumngrid
\appendix

\section{\label{Appendix::Sec::GZ_EigenState} GZ States as the Eigenstates of the Spin-$S$ XYZ Hamiltonian}

\subsection{Proof}

The proof of the existence of the periodic eigenstate in the XYZ Hamiltonian is provided by Granovskii and Zhedanov\,\cite{GranovskiiZhedanov1, GranovskiiZhedanov2}.
For the sake of completeness and the reader’s convenience, we reproduce the proof here.

First, we transform the Hamiltonian unitarily so that the GZ state becomes an all-up state in the rotated frame,
\begin{equation}
\ket{\Uparrow}=
    \prod_n \exp(i \opS{z}{n} \phi_n) \exp(i \opS{y}{n} \theta_n) \ket{\Psi_{GZ}},
\end{equation}
where $\ket{\Uparrow}$ denotes the product state of spins at each site with the quantum numbers, $\ket{S=S, S^z=S}$.
The expressions of $\phi_n$ and $\theta_n$ are given in Eq.\,\ref{Eq::GZScar}.
This unitary transformation corresponds to the following $SO(3)$ rotation of the spin operators,

\begin{equation}
    \begin{pmatrix}
        \opS{x}{n}\\
        \opS{y}{n}\\
        \opS{z}{n}
    \end{pmatrix}
    =
    \begin{pmatrix}
        -N_n\,\sn(w_n,\kappa) & -N_n \, \sn(\overline{w}_n,\kappa) & \alpha \,\cn(nq,\kappa)
        \\
        N_n \,\cn(w_n,\kappa) & N_n \,\cn(\overline{w}_n,\kappa) & \beta \, \sn(nq, \kappa)
        \\
        iN_n & -i N_n & \gamma\,\dn(nq,\kappa)
    \end{pmatrix}
    \begin{pmatrix}
        \ops{+}{n}
        \\
        \ops{-}{n}
        \\
        \ops{z}{n}
    \end{pmatrix},
\end{equation}
where $w_n=nq+iv$ and $v$ is the incomplete elliptic integral of the first kind with elliptic modulus $k^\prime=\sqrt{1-k^2}$ and $N_n=\sqrt{1-\gamma^2\,\dn^2(nq,\kappa)}$. 
$q$ is an arbitrary parameter, and in this section, we derive its relation to the exchange coefficients and show that the GZ state with a particular $q$-value is the eigenstate of the XYZ Hamiltonian.
XYZ Hamiltonian after the operator transformation becomes,
\begin{equation}
    \mathcal{H}^\prime
    =\sum_n 
    \left[
    \mathcal{C}_n^{--} \ops{-}{n} \ops{-}{n+1}
    +
    \mathcal{C}_n^{-+} \ops{-}{n} \ops{+}{n+1}
    +
    \mathcal{C}_n^{-z} \ops{-}{n} \ops{z}{n+1}
    +
    \mathcal{C}_n^{z-} \ops{z}{n} \ops{-}{n+1}
    +
    \text{H.C.}
    \right]
    +
    \mathcal{C}_n^{zz} \ops{z}{n} \ops{z}{n+1},
    \label{Appendix::Eq::RoatatedFrameHamiltonian}
\end{equation}
where the coefficients are given by,
\begin{gather}
    \mathcal{C}_n^{--}
    =
    N_n N_{n+1}\left[
    J_x\,\sn(\overline{w}_n,\kappa) \,\sn(\overline{w}_{n+1},\kappa)
    +
    J_y\,\cn(\overline{w}_n,\kappa) \,\cn(\overline{w}_{n+1},\kappa)
    -Jz
    \right]
    \nonumber\\
    \mathcal{C}_n^{-+}
    =
    N_nN_{n+1}\left[
    J_x\,\sn(\overline{w}_n)\,\sn(w_{n+1})
    +J_y\,\cn(\overline{w}_n) \,\cn(w_{n+1})
    +J_z
    \right]
    \nonumber\\
    \mathcal{C}_n^{-z}
    =
    -N_n\left[
    J_x\alpha\,\sn(\overline{w}_n) \,\cn(nq+q)
    -
    J_y \beta \,\cn(\overline{w}_n) \,\sn(nq+q)
    +
    iJ_z\gamma \,\dn(nq+q)
    \right]
    \nonumber \\
    \mathcal{C}_n^{z-}
    =
    -N_{n+1} \left[
    J_x\alpha \,\cn(nq)\,\sn(\overline{w}_{n+1})
    -
    J_y \beta \,\sn(nq)\,\cn(\overline{w}_{n+1})+iJ_z\gamma\,\dn(nq)
    \right]
    \nonumber \\
    \mathcal{C}_n^{zz}
    =
    \left[
    J_x \alpha^2 \,\cn(nq)\,\cn(nq+q)
    +
    J_y \beta^2 \,\sn(nq) \,\sn(nq+q)
    +
    J_z \gamma^2 \,\dn(nq)\,\dn(nq+q)
    \right].
    \label{Appendix::Eq::AllCoefficients}
\end{gather}
If the state $\ket{\Psi}_{GZ}$ is an eigenstate of the XYZ Hamiltonian, then the following eigenvalue equation must be satisfied,
\begin{flalign}
    &\;\;\;\;\;\;\;\;\;\;\;
    \mathcal{H}^\prime \ket{\Uparrow}=E\ket{\Uparrow}&&
    \nonumber \\
    &\implies
    \sum_n
    \left[
    \left\lbrace
    \mathcal{C}_n^{--} \ops{-}{n} \ops{-}{n+1}
    +
    \mathcal{C}_n^{-+} \ops{-}{n} \ops{+}{n+1}
    +
    \mathcal{C}_n^{-z} \ops{-}{n} \ops{z}{n+1}
    +
    \mathcal{C}_n^{z-} \ops{z}{n} \ops{-}{n+1}
    +
    \text{H.C.}
    \right\rbrace
    +
    \mathcal{C}_n^{zz} \ops{z}{n} \ops{z}{n+1}
    \right]
    \ket{\Uparrow}
    = E\ket{\Uparrow}&&
    \nonumber \\
    &\implies
    \sum_n
    \left[
    \mathcal{C}_n^{--} 
    \ops{-}{n}
    \ops{-}{n+1}
    +
    \mathcal{C}_{n}^{-z} 
    \ops{-}{n}
    \ops{z}{n+1}
    +
    \mathcal{C}_{n}^{z-}
    \ops{z}{n}
    \ops{-}{n+1}
    +
    \mathcal{C}_n^{zz}
    \ops{z}{n}
    \ops{z}{n+1}
    \right]
    \ket{\Uparrow}
    =E\ket{\Uparrow}&&
    \nonumber \\
    &\implies
    \sum_n
    \left[
    \mathcal{C}_n^{--} 
    \ops{-}{n}
    \ops{-}{n+1}
    +S
    \mathcal{C}_{n}^{-z} 
    \ops{-}{n}
    +S
    \mathcal{C}_{n}^{z-}
    \ops{-}{n+1}
    +S^2
    \mathcal{C}_n^{zz}
    \right]
    \ket{\Uparrow}
    =E\ket{\Uparrow}&&
    \nonumber \\
    &\implies
    \sum_n
    \left[
    \mathcal{C}_n^{--} 
    \ops{-}{n}
    \ops{-}{n+1}
    +S
    \left(
    \mathcal{C}_{n}^{-z} 
    +
    \mathcal{C}_{n-1}^{z-}
    \right)
    \ops{-}{n}
    +S^2
    \mathcal{C}_n^{zz}
    \right]
    \ket{\Uparrow}
    =E\ket{\Uparrow}.&&
\end{flalign}
For this equation to be a valid eigen equation, the following conditions must be satisfied,
\begin{equation}
    \mathcal{C}_n^{--}=0 
    \;\;\;\;\text{and}\;\;\;\;\;
    \mathcal{C}_{n}^{-z} 
    +
    \mathcal{C}_{n-1}^{z-}=0.
\end{equation}
This condition holds if,
\begin{equation}
    \dn(q,\kappa)=\frac{J_x}{J_y},\, \cn(q,\kappa)=\frac{J_z}{J_y},\,
\end{equation}
and it can be proved by using the following identities,
\begin{gather}
\cn(a,\kappa)\,\cn(b,\kappa)+\dn(a-b,\kappa)\sn(a,\kappa)\sn(b,\kappa)-\cn(a-b,\kappa)=0
    \nonumber \\
    \dn(a-b,\kappa)\sn(a,\kappa)\cn(b,\kappa)-
    \cn(a,\kappa)\sn(b,\kappa)
    -\dn(a,\kappa)\sn(a-b,\kappa)=0.
\end{gather}

\subsection{Explicit form of the GZ states}

Here we express the GZ state in the local $S^z$-basis for different spin values.
For spin values $S=1/2$, $1$ and $3/2$, the GZ states, respectively,
\begin{equation*}
\begin{split}
    \ket{\Psi_{\text{GZ}}^{S=1/2}}
    &=
    \prod_n \left[
    e^{-i\frac{\phi_n}{2}} \cos\left(\frac{\theta_n}{2}\right)
    \ket{S^z=1/2}_n
    + e^{i\frac{\phi_n}{2}} \sin\left(\frac{\theta_n}{2}\right)
    \ket{S^z=-1/2}_n
    \right]
    \\
    \ket{\Psi_{\text{GZ}}^{S=1}}
    &=
    \prod_n \left[
    e^{-i\frac{\phi_n}{2}} \cos^2\left(\frac{\theta_n}{2}\right)
    \ket{S^z=1}_n
    +\sqrt{2} \sin\left(\frac{\theta_n}{2}\right) \cos\left(\frac{\theta_n}{2}\right)
    \ket{S^z=0}_n
    + e^{i\frac{\phi_n}{2}} \sin^2\left(\frac{\theta_n}{2}\right)
    \ket{S^z=-1}_n
    \right]
    \\
    \ket{\Psi_{\text{GZ}}^{S=3/2}}
    &=
    \prod_n \left[
    e^{-i\frac{3\phi_n}{2}} \cos^3\left(\frac{\theta_n}{2}\right)
    \ket{S^z=3/2}_n
    +\sqrt{3} e^{-i\frac{\phi_n}{2}} \sin\left(\frac{\theta_n}{2}\right) \cos^2\left(\frac{\theta_n}{2}\right)
    \ket{S^z=1/2}_n
    \right.
    \\
    &\qquad\qquad\qquad\left.
    +\sqrt{3} e^{i\frac{\phi_n}{2}} \sin^2\left(\frac{\theta_n}{2}\right) \cos\left(\frac{\theta_n}{2}\right)
    \ket{S^z=-1/2}_n
    + e^{i\frac{3\phi_n}{2}} \sin^3\left(\frac{\theta_n}{2}\right)
    \ket{S^z=-3/2}_n
    \right]
\end{split}
\end{equation*}

\section{\label{Appendix::Sec::AlternativeStabilizers}  Additional Numerical Results}

\begin{figure}[t]
\centering
\includegraphics[width=1.0\textwidth]{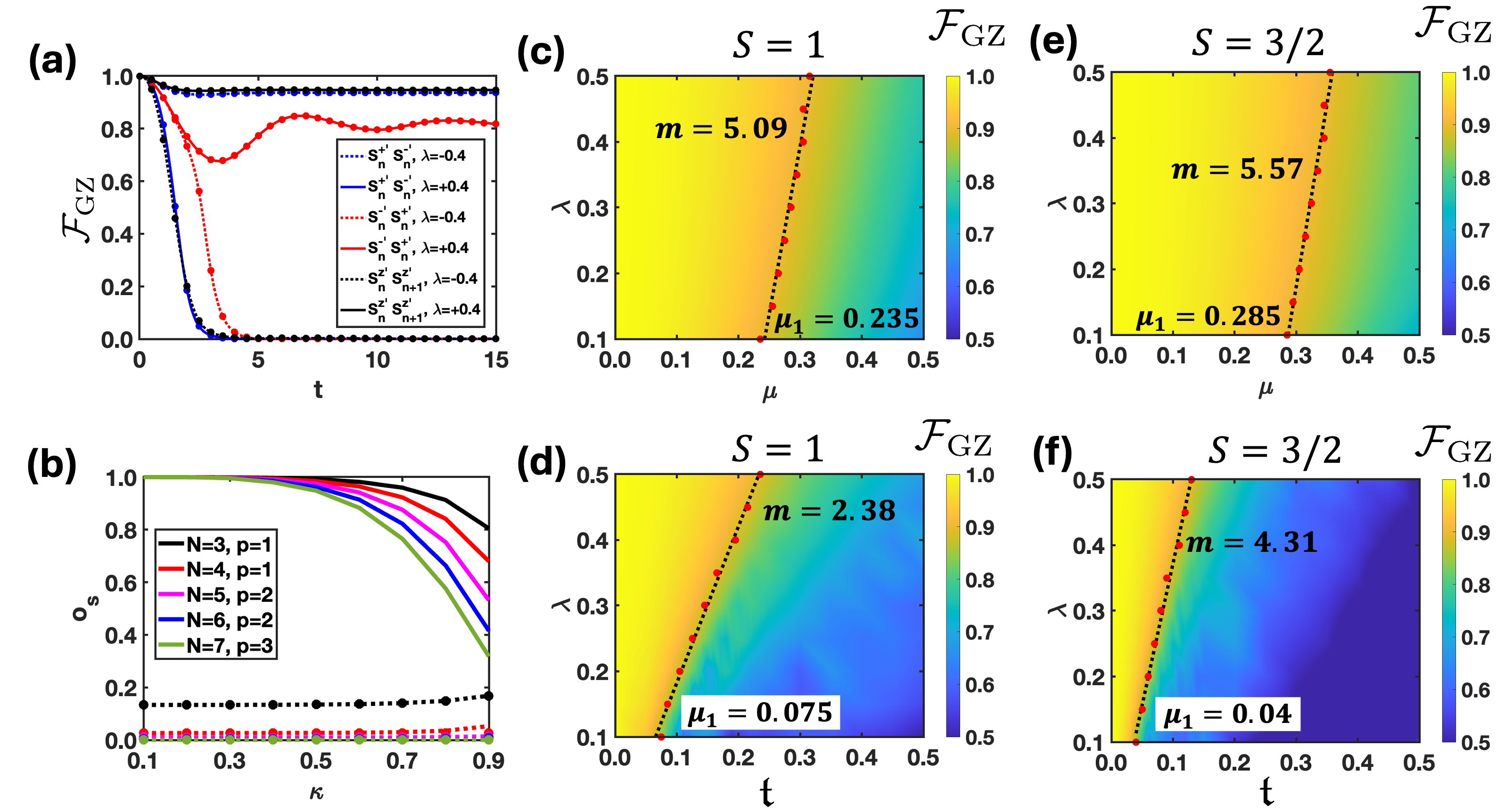}
\caption{
(a)
Stabilization of GZ state using alternative non-Hermitian terms in the presence of global onsite perturbation $\mu=0.3$.
The simulations are done for parameters $\kappa=0.5$, $\gamma=0.8$, $q=4K(\kappa)/5$.
(b)
The overlap $o_s$ of the spin-$1$ states $\ket{1111\ldots}$, $\ket{1\bar{1}1\bar{1}\ldots}$, and $\ket{1\bar{1}\bar{1}1\bar{1}\bar{1}\ldots}$ with the GZ scar subspace for finite periodic chain of length $N$, is shown as a function of $\kappa$, represented by solid lines, dashed lines, and dots, respectively.
The different colored plots correspond to the GZ scar subspace for different values of $q$, which is tuned by varying the parameters $N$ and $p$, since $q = \frac{4pK(\kappa)}{N}$.
The phase diagram representing stabilization of the GZ scar states for the (c) parameter space $(\mu,\lambda)$ and spin $S=1$, (d) parameter space $(\mathfrak{t},\lambda)$ and spin $S=1$, (e) parameter space $(\mu,\lambda)$ and spin $S=3/2$, (f) parameter space $(\mathfrak{t},\lambda)$ and spin $S=3/2$.
The colormap denotes the fidelity of the time-evolved state at $t=15$ with respect to the initial GZ state, where $t=15$ corresponds to the steady state with saturated fidelity.
The boundary line (black dashed line) is plotted via fitting the boundary points (red dots) that separate the fidelity regimes $\mathcal{F}_{\text{GZ}}<0.9$ and $\mathcal{F}_{\text{GZ}}>0.9$.
The parameters $m$ and $\mu_1$ represent the slope of the boundary line and the $\mu$-value of the point at which it intersects the line $\lambda = 0.1$, respectively. 
The system parameters used for the simulation for the plots (c)-(f) are $\kappa = 0.5$, $\gamma = 0.8$, $q = 4K(\kappa)/5$, $h=0.0$.
Additionally, we use iTEBD parameter set $(\delta\tau = 5 \times 10^{-4}, \chi = 140)$ for the simulation.
}
\label{fig::Apendix::ExtraRelaxation}
\end{figure}

\subsection{Alternative non-Hermitian Stabilizing Terms}

In the main text, we show the stabilization of the GZ state using one kind of non-Hermitian term, which has a physical interpretation as spin relaxation.
However, mathematically, there are infinitely many possible non-Hermitian constructs that are capable of stabilizing the GZ state, such as,
\begin{equation}
    i\lambda\sum_n \opS{+^\prime}{n} \opS{-^\prime}{n},\quad
    i\lambda\sum_n \opS{-^\prime}{n} \opS{+^\prime}{n},\quad
    i\lambda\sum_n \opS{z^\prime}{n} \opS{z^\prime}{n+1}.
\end{equation}
The GZ states are eigenstate of the operators $\opS{+^\prime}{n} \opS{-^\prime}{n}$, $\opS{-^\prime}{n} \opS{+^\prime}{n}$, $\opS{z^\prime}{n} \opS{z^\prime}{n+1}$.
The Fig.\,\ref{fig::Apendix::ExtraRelaxation}(a) shows the stabilization of the GZ state for positive $\lambda$ values for the operators $\opS{-^\prime}{n} \opS{+^\prime}{n}$, $\opS{z^\prime}{n} \opS{z^\prime}{n+1}$ and negative $\lambda$ values of $\opS{z^\prime}{n} \opS{z^\prime}{n+1}$.
The simulations are done using iTEBD in the presence of a global onsite perturbation.

\subsection{Overlap of the Product States with the GZ Scar Subspace}

We quantify the overlap between the product states $\ket{s}$ with the GZ scar subspace for a finite periodic system, as follows,
\begin{equation}
    o_s = \sum_i \left|\left\langle s\middle|i_{\text{GZ}}\right\rangle\right|^2,
\end{equation}
where $i_{\text{GZ}}$ denotes the $i$-th orthonormal state in the GZ subspace, and the product state $\ket{s}$ is chosen from spin-$1$ configurations such as $\ket{1111\ldots}$, $\ket{1\bar{1}1\bar{1}\ldots}$, or $\ket{1\bar{1}\bar{1}1\bar{1}\bar{1}\ldots}$.
Next we discuss the procedure to generate orthonormal states in the GZ scar subspace.
Firstly, we generate $4NS$ non-orthonormal states as follows,
\begin{equation}
    \ket{\phi_1}=\hat{\mathcal{R}}(\gamma=0,\kappa,\phi=0)\ket{\Uparrow},\,
    \ket{\phi_2}=\hat{\mathcal{R}}(\gamma=0,\kappa,\phi=\pi/M)\ket{\Uparrow},\,
    \ket{\phi_3}=\hat{\mathcal{R}}(\gamma=0,\kappa,\phi=2\pi/M)\ket{\Uparrow},\,
    \ldots,
\end{equation}
where $M=2NS$.
Moreover the rotation operator $\hat{\mathcal{R}}(\gamma,\kappa,\phi)$ is defined as $\hat{\mathcal{R}}(\gamma,\kappa,\phi)=\exp(- i \opS{z}{n} \phi_n) \exp(-i \opS{y}{n} \theta_n)$ where the variables $(\theta_n, \phi_n)$ are functions of $\gamma$, $\kappa$, and $\phi$, as defined in Eq.\,\ref{Eq::GZScar} of the main text.
From this non-orthonormal set of states, we generate the $4NS$ orthonormal set of states using  Gram–Schmidt orthonormalization protocol,
\begin{equation}
    \ket{1}=\ket{\phi_1},\,
    \ket{2}=\ket{\phi_2}-\braket{1}{\phi_2} \ket{1},\,
    \ket{3}=\ket{\phi_3}
    -\braket{1}{\phi_3} \ket{1}
    -\braket{2}{\phi_3} \ket{2},
    \cdots(4NS\text{-times}).
\end{equation}
In Fig.\,\ref{fig::Apendix::ExtraRelaxation}(b), we plot the overlap $o_s$ as a function of $\kappa$ for the different states  $\ket{1111\ldots}$, $\ket{1\bar{1}1\bar{1}\ldots}$, or $\ket{1\bar{1}\bar{1}1\bar{1}\bar{1}\ldots}$ with the GZ scar subspace and for different $q$-values determined by the parameters $N$ and $p$ as $q=4pK(\kappa)/N$.
We observe that the GZ scar subspace exhibits a high overlap with the product state $\ket{1111\ldots}$, which decreases as a function of $\kappa$. 
As a consequence, the non-Hermitian relaxation protocol is highly effective in stabilizing the state $\ket{1111\ldots}$, as shown in Fig.\,\ref{fig::Dynamics}(e).
In contrast, the states $\ket{1\bar{1}1\bar{1}\ldots}$ and $\ket{1\bar{1}\bar{1}1\bar{1}\bar{1}\ldots}$ have a much smaller overlap with the GZ scar subspace and therefore cannot be stabilized under the same protocol, as also illustrated in Fig.\,\ref{fig::Dynamics}(e).

\subsection{Stability Phase Diagram}

We plot the stability phase diagram in Fig.\,\ref{fig::Apendix::ExtraRelaxation}(c)–(f), where the colormap represents the fidelity of the state at time $t$ with respect to the GZ state at $t=15$ for an infinite system. 
The choice of $t=15$ corresponds to the steady-state regime, as shown in the main text.
The phase diagram is presented for two parameter spaces: $(\mu, \lambda)$ and $(\mathfrak{t}, \lambda)$.
To further characterize the phase diagram, we introduce a boundary that delineates regions with fidelity $\mathcal{F}_{\text{GZ}} > 0.9$ and $\mathcal{F}_{\text{GZ}} < 0.9$.

In Fig.\,\ref{fig::Apendix::ExtraRelaxation}(c) and (e), we present the stability phase diagram in the $(\mu,\lambda)$ parameter space for spin-$1$ and spin-$3/2$ systems, respectively.
As the spin increases from $S=1$ to $S=3/2$, the boundary between the fidelity regimes $\mathcal{F}_{\text{GZ}} > 0.9$ and $\mathcal{F}_{\text{GZ}} < 0.9$ is observed to shift toward higher $\mu$ values, accompanied by a slight increase in its slope.
These observations indicate that, with increasing spin, the GZ scar states exhibit enhanced stability against onsite perturbations.

In contrast, for the hopping perturbation, we observe the opposite, meaning higher spin values have lesser stability.
In Fig.\,\ref{fig::Apendix::ExtraRelaxation}(d) and (f), we present the stability phase diagram in the $(\mathfrak{t},\lambda)$ parameter space for spin-$1$ and spin-$3/2$ systems, respectively.
As the spin increases from $S=1$ to $S=3/2$, the boundary between the fidelity regimes $\mathcal{F}_{\text{GZ}} > 0.9$ and $\mathcal{F}_{\text{GZ}} < 0.9$ is observed to shift toward lower $\mathfrak{t}$ values, accompanied by a large increase in its slope.
These observations indicate that, with increasing spin, the GZ scar states exhibit diminished stability against hopping perturbations.

Notably, the stark contrast in the stability phase diagrams for onsite and hopping perturbations can be partly attributed to their local and non-local nature, respectively, while the non-Hermitian spin relaxation term remains local in nature.

\twocolumngrid


\begin{thebibliography}{76}%
\makeatletter
\providecommand \@ifxundefined [1]{%
 \@ifx{#1\undefined}
}%
\providecommand \@ifnum [1]{%
 \ifnum #1\expandafter \@firstoftwo
 \else \expandafter \@secondoftwo
 \fi
}%
\providecommand \@ifx [1]{%
 \ifx #1\expandafter \@firstoftwo
 \else \expandafter \@secondoftwo
 \fi
}%
\providecommand \natexlab [1]{#1}%
\providecommand \enquote  [1]{``#1''}%
\providecommand \bibnamefont  [1]{#1}%
\providecommand \bibfnamefont [1]{#1}%
\providecommand \citenamefont [1]{#1}%
\providecommand \href@noop [0]{\@secondoftwo}%
\providecommand \href [0]{\begingroup \@sanitize@url \@href}%
\providecommand \@href[1]{\@@startlink{#1}\@@href}%
\providecommand \@@href[1]{\endgroup#1\@@endlink}%
\providecommand \@sanitize@url [0]{\catcode `\\12\catcode `\$12\catcode `\&12\catcode `\#12\catcode `\^12\catcode `\_12\catcode `\%12\relax}%
\providecommand \@@startlink[1]{}%
\providecommand \@@endlink[0]{}%
\providecommand \url  [0]{\begingroup\@sanitize@url \@url }%
\providecommand \@url [1]{\endgroup\@href {#1}{\urlprefix }}%
\providecommand \urlprefix  [0]{URL }%
\providecommand \Eprint [0]{\href }%
\providecommand \doibase [0]{http://dx.doi.org/}%
\providecommand \selectlanguage [0]{\@gobble}%
\providecommand \bibinfo  [0]{\@secondoftwo}%
\providecommand \bibfield  [0]{\@secondoftwo}%
\providecommand \translation [1]{[#1]}%
\providecommand \BibitemOpen [0]{}%
\providecommand \bibitemStop [0]{}%
\providecommand \bibitemNoStop [0]{.\EOS\space}%
\providecommand \EOS [0]{\spacefactor3000\relax}%
\providecommand \BibitemShut  [1]{\csname bibitem#1\endcsname}%
\let\auto@bib@innerbib\@empty
\bibitem [{\citenamefont {Neumann}(1929)}]{ETH1}%
  \BibitemOpen
  \bibfield  {author} {\bibinfo {author} {\bibfnamefont {J.~v.}\ \bibnamefont {Neumann}},\ }\bibfield  {title} {\textit {\bibinfo {title} {Beweis des Ergodensatzes und desH-Theorems in der neuen Mechanik},\ }}\href {\doibase 10.1007/BF01339852} {\bibfield  {journal} {\bibinfo  {journal} {Z. Physik}\ }\textbf {\bibinfo {volume} {57}},\ \bibinfo {pages} {30} (\bibinfo {year} {1929})}\BibitemShut {NoStop}%
\bibitem [{\citenamefont {Neumann}(2010)}]{ETH2}%
  \BibitemOpen
  \bibfield  {author} {\bibinfo {author} {\bibfnamefont {J.~v.}\ \bibnamefont {Neumann}},\ }\bibfield  {title} {\textit {\bibinfo {title} {Proof of the ergodic theorem and the H-theorem in quantum mechanics},\ }}\href {\doibase 10.1140/epjh/e2010-00008-5} {\bibfield  {journal} {\bibinfo  {journal} {Z. Physik}\ }\textbf {\bibinfo {volume} {35}},\ \bibinfo {pages} {201} (\bibinfo {year} {2010})}\BibitemShut {NoStop}%
\bibitem [{\citenamefont {Deutsch}(1991)}]{ETH3}%
  \BibitemOpen
  \bibfield  {author} {\bibinfo {author} {\bibfnamefont {J.~M.}\ \bibnamefont {Deutsch}},\ }\bibfield  {title} {\textit {\bibinfo {title} {Quantum statistical mechanics in a closed system},\ }}\href {\doibase 10.1103/PhysRevA.43.2046} {\bibfield  {journal} {\bibinfo  {journal} {Phys. Rev. A}\ }\textbf {\bibinfo {volume} {43}},\ \bibinfo {pages} {2046} (\bibinfo {year} {1991})}\BibitemShut {NoStop}%
\bibitem [{\citenamefont {Srednicki}(1994)}]{ETH4}%
  \BibitemOpen
  \bibfield  {author} {\bibinfo {author} {\bibfnamefont {M.}~\bibnamefont {Srednicki}},\ }\bibfield  {title} {\textit {\bibinfo {title} {Chaos and quantum thermalization},\ }}\href {\doibase 10.1103/PhysRevE.50.888} {\bibfield  {journal} {\bibinfo  {journal} {Phys. Rev. E}\ }\textbf {\bibinfo {volume} {50}},\ \bibinfo {pages} {888} (\bibinfo {year} {1994})}\BibitemShut {NoStop}%
\bibitem [{\citenamefont {Srednicki}(1996)}]{ETH_more1}%
  \BibitemOpen
  \bibfield  {author} {\bibinfo {author} {\bibfnamefont {M.}~\bibnamefont {Srednicki}},\ }\bibfield  {title} {\textit {\bibinfo {title} {Thermal fluctuations in quantized chaotic systems},\ }}\href {\doibase 10.1088/0305-4470/29/4/003} {\bibfield  {journal} {\bibinfo  {journal} {Journal of Physics A: Mathematical and General}\ }\textbf {\bibinfo {volume} {29}},\ \bibinfo {pages} {L75} (\bibinfo {year} {1996})}\BibitemShut {NoStop}%
\bibitem [{\citenamefont {Rigol}\ \emph {et~al.}(2008)\citenamefont {Rigol}, \citenamefont {Dunjko},\ and\ \citenamefont {Olshanii}}]{ETH_more2}%
  \BibitemOpen
  \bibfield  {author} {\bibinfo {author} {\bibfnamefont {M.}~\bibnamefont {Rigol}}, \bibinfo {author} {\bibfnamefont {V.}~\bibnamefont {Dunjko}}, \ and\ \bibinfo {author} {\bibfnamefont {M.}~\bibnamefont {Olshanii}},\ }\bibfield  {title} {\textit {\bibinfo {title} {Thermalization and its mechanism for generic isolated quantum systems},\ }}\href {\doibase 10.1038/nature06838} {\bibfield  {journal} {\bibinfo  {journal} {Nature}\ }\textbf {\bibinfo {volume} {452}},\ \bibinfo {pages} {854} (\bibinfo {year} {2008})}\BibitemShut {NoStop}%
\bibitem [{\citenamefont {D'Alessio}\ \emph {et~al.}(2016)\citenamefont {D'Alessio}, \citenamefont {Kafri}, \citenamefont {Polkovnikov},\ and\ \citenamefont {Rigol}}]{ETH_more3}%
  \BibitemOpen
  \bibfield  {author} {\bibinfo {author} {\bibfnamefont {L.}~\bibnamefont {D'Alessio}}, \bibinfo {author} {\bibfnamefont {Y.}~\bibnamefont {Kafri}}, \bibinfo {author} {\bibfnamefont {A.}~\bibnamefont {Polkovnikov}}, \ and\ \bibinfo {author} {\bibfnamefont {M.}~\bibnamefont {Rigol}},\ }\bibfield  {title} {\textit {\bibinfo {title} {From quantum chaos and eigenstate thermalization to statistical mechanics and thermodynamics},\ }}\href {\doibase 10.1080/00018732.2016.1198134} {\bibfield  {journal} {\bibinfo  {journal} {Advances in Physics}\ }\textbf {\bibinfo {volume} {65}},\ \bibinfo {pages} {239} (\bibinfo {year} {2016})}\BibitemShut {NoStop}%
\bibitem [{\citenamefont {Deutsch}(2018)}]{ETH_more4}%
  \BibitemOpen
  \bibfield  {author} {\bibinfo {author} {\bibfnamefont {J.~M.}\ \bibnamefont {Deutsch}},\ }\bibfield  {title} {\textit {\bibinfo {title} {Eigenstate thermalization hypothesis},\ }}\href {\doibase 10.1088/1361-6633/aac9f1} {\bibfield  {journal} {\bibinfo  {journal} {Rep. Prog. Phys}\ }\textbf {\bibinfo {volume} {81}},\ \bibinfo {pages} {082001} (\bibinfo {year} {2018})}\BibitemShut {NoStop}%
\bibitem [{\citenamefont {Srednicki}(1999)}]{ETH_more5}%
  \BibitemOpen
  \bibfield  {author} {\bibinfo {author} {\bibfnamefont {M.}~\bibnamefont {Srednicki}},\ }\bibfield  {title} {\textit {\bibinfo {title} {The approach to thermal equilibrium in quantized chaotic systems},\ }}\href {\doibase 10.1088/0305-4470/32/7/007} {\bibfield  {journal} {\bibinfo  {journal} {Journal of Physics A: Mathematical and General}\ }\textbf {\bibinfo {volume} {32}},\ \bibinfo {pages} {1163} (\bibinfo {year} {1999})}\BibitemShut {NoStop}%
\bibitem [{\citenamefont {Nandkishore}\ and\ \citenamefont {Huse}(2015)}]{MBL1}%
  \BibitemOpen
  \bibfield  {author} {\bibinfo {author} {\bibfnamefont {R.}~\bibnamefont {Nandkishore}}\ and\ \bibinfo {author} {\bibfnamefont {D.~A.}\ \bibnamefont {Huse}},\ }\bibfield  {title} {\textit {\bibinfo {title} {Many-Body Localization and Thermalization in Quantum Statistical Mechanics},\ }}\href {\doibase https://doi.org/10.1146/annurev-conmatphys-031214-014726} {\bibfield  {journal} {\bibinfo  {journal} {Annual Review of Condensed Matter Physics}\ }\textbf {\bibinfo {volume} {6}},\ \bibinfo {pages} {15} (\bibinfo {year} {2015})}\BibitemShut {NoStop}%
\bibitem [{\citenamefont {Abanin}\ \emph {et~al.}(2019)\citenamefont {Abanin}, \citenamefont {Altman}, \citenamefont {Bloch},\ and\ \citenamefont {Serbyn}}]{MBL2}%
  \BibitemOpen
  \bibfield  {author} {\bibinfo {author} {\bibfnamefont {D.~A.}\ \bibnamefont {Abanin}}, \bibinfo {author} {\bibfnamefont {E.}~\bibnamefont {Altman}}, \bibinfo {author} {\bibfnamefont {I.}~\bibnamefont {Bloch}}, \ and\ \bibinfo {author} {\bibfnamefont {M.}~\bibnamefont {Serbyn}},\ }\bibfield  {title} {\textit {\bibinfo {title} {Colloquium: Many-body localization, thermalization, and entanglement},\ }}\href {\doibase 10.1103/RevModPhys.91.021001} {\bibfield  {journal} {\bibinfo  {journal} {Rev. Mod. Phys.}\ }\textbf {\bibinfo {volume} {91}},\ \bibinfo {pages} {021001} (\bibinfo {year} {2019})}\BibitemShut {NoStop}%
\bibitem [{\citenamefont {Turner}\ \emph {et~al.}(2018{\natexlab{a}})\citenamefont {Turner}, \citenamefont {Michailidis}, \citenamefont {Abanin}, \citenamefont {Serbyn},\ and\ \citenamefont {Papi\ifmmode~\acute{c}\else \'{c}\fi{}}}]{PXP1}%
  \BibitemOpen
  \bibfield  {author} {\bibinfo {author} {\bibfnamefont {C.~J.}\ \bibnamefont {Turner}}, \bibinfo {author} {\bibfnamefont {A.~A.}\ \bibnamefont {Michailidis}}, \bibinfo {author} {\bibfnamefont {D.~A.}\ \bibnamefont {Abanin}}, \bibinfo {author} {\bibfnamefont {M.}~\bibnamefont {Serbyn}}, \ and\ \bibinfo {author} {\bibfnamefont {Z.}~\bibnamefont {Papi\ifmmode~\acute{c}\else \'{c}\fi{}}},\ }\bibfield  {title} {\textit {\bibinfo {title} {Weak ergodicity breaking from quantum many-body scars},\ }}\href {\doibase 10.1038/s41567-018-0137-5} {\bibfield  {journal} {\bibinfo  {journal} {Nature Physics}\ }\textbf {\bibinfo {volume} {14}},\ \bibinfo {pages} {745} (\bibinfo {year} {2018}{\natexlab{a}})}\BibitemShut {NoStop}%
\bibitem [{\citenamefont {Turner}\ \emph {et~al.}(2018{\natexlab{b}})\citenamefont {Turner}, \citenamefont {Michailidis}, \citenamefont {Abanin}, \citenamefont {Serbyn},\ and\ \citenamefont {Papi\ifmmode~\acute{c}\else \'{c}\fi{}}}]{PXP2}%
  \BibitemOpen
  \bibfield  {author} {\bibinfo {author} {\bibfnamefont {C.~J.}\ \bibnamefont {Turner}}, \bibinfo {author} {\bibfnamefont {A.~A.}\ \bibnamefont {Michailidis}}, \bibinfo {author} {\bibfnamefont {D.~A.}\ \bibnamefont {Abanin}}, \bibinfo {author} {\bibfnamefont {M.}~\bibnamefont {Serbyn}}, \ and\ \bibinfo {author} {\bibfnamefont {Z.}~\bibnamefont {Papi\ifmmode~\acute{c}\else \'{c}\fi{}}},\ }\bibfield  {title} {\textit {\bibinfo {title} {Quantum scarred eigenstates in a Rydberg atom chain: Entanglement, breakdown of thermalization, and stability to perturbations},\ }}\href {\doibase 10.1103/PhysRevB.98.155134} {\bibfield  {journal} {\bibinfo  {journal} {Phys. Rev. B}\ }\textbf {\bibinfo {volume} {98}},\ \bibinfo {pages} {155134} (\bibinfo {year} {2018}{\natexlab{b}})}\BibitemShut {NoStop}%
\bibitem [{\citenamefont {James}\ \emph {et~al.}(2019)\citenamefont {James}, \citenamefont {Konik},\ and\ \citenamefont {Robinson}}]{Ising1}%
  \BibitemOpen
  \bibfield  {author} {\bibinfo {author} {\bibfnamefont {A.~J.~A.}\ \bibnamefont {James}}, \bibinfo {author} {\bibfnamefont {R.~M.}\ \bibnamefont {Konik}}, \ and\ \bibinfo {author} {\bibfnamefont {N.~J.}\ \bibnamefont {Robinson}},\ }\bibfield  {title} {\textit {\bibinfo {title} {Nonthermal States Arising from Confinement in One and Two Dimensions},\ }}\href {\doibase 10.1103/PhysRevLett.122.130603} {\bibfield  {journal} {\bibinfo  {journal} {Phys. Rev. Lett.}\ }\textbf {\bibinfo {volume} {122}},\ \bibinfo {pages} {130603} (\bibinfo {year} {2019})}\BibitemShut {NoStop}%
\bibitem [{\citenamefont {Robinson}\ \emph {et~al.}(2019)\citenamefont {Robinson}, \citenamefont {James},\ and\ \citenamefont {Konik}}]{Ising2}%
  \BibitemOpen
  \bibfield  {author} {\bibinfo {author} {\bibfnamefont {N.~J.}\ \bibnamefont {Robinson}}, \bibinfo {author} {\bibfnamefont {A.~J.~A.}\ \bibnamefont {James}}, \ and\ \bibinfo {author} {\bibfnamefont {R.~M.}\ \bibnamefont {Konik}},\ }\bibfield  {title} {\textit {\bibinfo {title} {Signatures of rare states and thermalization in a theory with confinement},\ }}\href {\doibase 10.1103/PhysRevB.99.195108} {\bibfield  {journal} {\bibinfo  {journal} {Phys. Rev. B}\ }\textbf {\bibinfo {volume} {99}},\ \bibinfo {pages} {195108} (\bibinfo {year} {2019})}\BibitemShut {NoStop}%
\bibitem [{\citenamefont {Vafek}\ \emph {et~al.}(2017)\citenamefont {Vafek}, \citenamefont {Regnault},\ and\ \citenamefont {Bernevig}}]{Hubbard1}%
  \BibitemOpen
  \bibfield  {author} {\bibinfo {author} {\bibfnamefont {O.}~\bibnamefont {Vafek}}, \bibinfo {author} {\bibfnamefont {N.}~\bibnamefont {Regnault}}, \ and\ \bibinfo {author} {\bibfnamefont {B.~A.}\ \bibnamefont {Bernevig}},\ }\bibfield  {title} {\textit {\bibinfo {title} {{Entanglement of exact excited eigenstates of the Hubbard model in arbitrary dimension}},\ }}\href {\doibase 10.21468/SciPostPhys.3.6.043} {\bibfield  {journal} {\bibinfo  {journal} {SciPost Phys.}\ }\textbf {\bibinfo {volume} {3}},\ \bibinfo {pages} {043} (\bibinfo {year} {2017})}\BibitemShut {NoStop}%
\bibitem [{\citenamefont {Iadecola}\ and\ \citenamefont {\ifmmode \check{Z}\else \v{Z}\fi{}nidari\ifmmode~\check{c}\else \v{c}\fi{}}(2019)}]{Hubbard2}%
  \BibitemOpen
  \bibfield  {author} {\bibinfo {author} {\bibfnamefont {T.}~\bibnamefont {Iadecola}}\ and\ \bibinfo {author} {\bibfnamefont {M.}~\bibnamefont {\ifmmode \check{Z}\else \v{Z}\fi{}nidari\ifmmode~\check{c}\else \v{c}\fi{}}},\ }\bibfield  {title} {\textit {\bibinfo {title} {Exact Localized and Ballistic Eigenstates in Disordered Chaotic Spin Ladders and the Fermi-Hubbard Model},\ }}\href {\doibase 10.1103/PhysRevLett.123.036403} {\bibfield  {journal} {\bibinfo  {journal} {Phys. Rev. Lett.}\ }\textbf {\bibinfo {volume} {123}},\ \bibinfo {pages} {036403} (\bibinfo {year} {2019})}\BibitemShut {NoStop}%
\bibitem [{\citenamefont {Mark}\ and\ \citenamefont {Motrunich}(2020)}]{Hubbard3}%
  \BibitemOpen
  \bibfield  {author} {\bibinfo {author} {\bibfnamefont {D.~K.}\ \bibnamefont {Mark}}\ and\ \bibinfo {author} {\bibfnamefont {O.~I.}\ \bibnamefont {Motrunich}},\ }\bibfield  {title} {\textit {\bibinfo {title} {$\ensuremath{\eta}$-pairing states as true scars in an extended Hubbard model},\ }}\href {\doibase 10.1103/PhysRevB.102.075132} {\bibfield  {journal} {\bibinfo  {journal} {Phys. Rev. B}\ }\textbf {\bibinfo {volume} {102}},\ \bibinfo {pages} {075132} (\bibinfo {year} {2020})}\BibitemShut {NoStop}%
\bibitem [{\citenamefont {Moudgalya}\ \emph {et~al.}(2020{\natexlab{a}})\citenamefont {Moudgalya}, \citenamefont {Regnault},\ and\ \citenamefont {Bernevig}}]{Hubbard4}%
  \BibitemOpen
  \bibfield  {author} {\bibinfo {author} {\bibfnamefont {S.}~\bibnamefont {Moudgalya}}, \bibinfo {author} {\bibfnamefont {N.}~\bibnamefont {Regnault}}, \ and\ \bibinfo {author} {\bibfnamefont {B.~A.}\ \bibnamefont {Bernevig}},\ }\bibfield  {title} {\textit {\bibinfo {title} {$\ensuremath{\eta}$-pairing in Hubbard models: From spectrum generating algebras to quantum many-body scars},\ }}\href {\doibase 10.1103/PhysRevB.102.085140} {\bibfield  {journal} {\bibinfo  {journal} {Phys. Rev. B}\ }\textbf {\bibinfo {volume} {102}},\ \bibinfo {pages} {085140} (\bibinfo {year} {2020}{\natexlab{a}})}\BibitemShut {NoStop}%
\bibitem [{\citenamefont {Moudgalya}\ \emph {et~al.}(2020{\natexlab{b}})\citenamefont {Moudgalya}, \citenamefont {Bernevig},\ and\ \citenamefont {Regnault}}]{Hall1}%
  \BibitemOpen
  \bibfield  {author} {\bibinfo {author} {\bibfnamefont {S.}~\bibnamefont {Moudgalya}}, \bibinfo {author} {\bibfnamefont {B.~A.}\ \bibnamefont {Bernevig}}, \ and\ \bibinfo {author} {\bibfnamefont {N.}~\bibnamefont {Regnault}},\ }\bibfield  {title} {\textit {\bibinfo {title} {Quantum many-body scars in a Landau level on a thin torus},\ }}\href {\doibase 10.1103/PhysRevB.102.195150} {\bibfield  {journal} {\bibinfo  {journal} {Phys. Rev. B}\ }\textbf {\bibinfo {volume} {102}},\ \bibinfo {pages} {195150} (\bibinfo {year} {2020}{\natexlab{b}})}\BibitemShut {NoStop}%
\bibitem [{\citenamefont {Nachtergaele}\ \emph {et~al.}(2020)\citenamefont {Nachtergaele}, \citenamefont {Warzel},\ and\ \citenamefont {Young}}]{Hall2}%
  \BibitemOpen
  \bibfield  {author} {\bibinfo {author} {\bibfnamefont {B.}~\bibnamefont {Nachtergaele}}, \bibinfo {author} {\bibfnamefont {S.}~\bibnamefont {Warzel}}, \ and\ \bibinfo {author} {\bibfnamefont {A.}~\bibnamefont {Young}},\ }\bibfield  {title} {\textit {\bibinfo {title} {Low-complexity eigenstates of a $\nu$ = 1/3 fractional quantum Hall system},\ }}\href {\doibase 10.1088/1751-8121/abca73} {\bibfield  {journal} {\bibinfo  {journal} {Journal of Physics A: Mathematical and Theoretical}\ }\textbf {\bibinfo {volume} {54}},\ \bibinfo {pages} {01LT01} (\bibinfo {year} {2020})}\BibitemShut {NoStop}%
\bibitem [{\citenamefont {Pai}\ and\ \citenamefont {Pretko}(2019)}]{Fracton1}%
  \BibitemOpen
  \bibfield  {author} {\bibinfo {author} {\bibfnamefont {S.}~\bibnamefont {Pai}}\ and\ \bibinfo {author} {\bibfnamefont {M.}~\bibnamefont {Pretko}},\ }\bibfield  {title} {\textit {\bibinfo {title} {Dynamical Scar States in Driven Fracton Systems},\ }}\href {\doibase 10.1103/PhysRevLett.123.136401} {\bibfield  {journal} {\bibinfo  {journal} {Phys. Rev. Lett.}\ }\textbf {\bibinfo {volume} {123}},\ \bibinfo {pages} {136401} (\bibinfo {year} {2019})}\BibitemShut {NoStop}%
\bibitem [{\citenamefont {Ok}\ \emph {et~al.}(2019)\citenamefont {Ok}, \citenamefont {Choo}, \citenamefont {Mudry}, \citenamefont {Castelnovo}, \citenamefont {Chamon},\ and\ \citenamefont {Neupert}}]{Fracton2}%
  \BibitemOpen
  \bibfield  {author} {\bibinfo {author} {\bibfnamefont {S.}~\bibnamefont {Ok}}, \bibinfo {author} {\bibfnamefont {K.}~\bibnamefont {Choo}}, \bibinfo {author} {\bibfnamefont {C.}~\bibnamefont {Mudry}}, \bibinfo {author} {\bibfnamefont {C.}~\bibnamefont {Castelnovo}}, \bibinfo {author} {\bibfnamefont {C.}~\bibnamefont {Chamon}}, \ and\ \bibinfo {author} {\bibfnamefont {T.}~\bibnamefont {Neupert}},\ }\bibfield  {title} {\textit {\bibinfo {title} {Topological many-body scar states in dimensions one, two, and three},\ }}\href {\doibase 10.1103/PhysRevResearch.1.033144} {\bibfield  {journal} {\bibinfo  {journal} {Phys. Rev. Res.}\ }\textbf {\bibinfo {volume} {1}},\ \bibinfo {pages} {033144} (\bibinfo {year} {2019})}\BibitemShut {NoStop}%
\bibitem [{\citenamefont {Shiraishi}(2019)}]{AKLT1}%
  \BibitemOpen
  \bibfield  {author} {\bibinfo {author} {\bibfnamefont {N.}~\bibnamefont {Shiraishi}},\ }\bibfield  {title} {\textit {\bibinfo {title} {Connection between quantum-many-body scars and the Affleck–Kennedy–Lieb–Tasaki model from the viewpoint of embedded Hamiltonians},\ }}\href {\doibase 10.1088/1742-5468/ab342e} {\bibfield  {journal} {\bibinfo  {journal} {Journal of Statistical Mechanics: Theory and Experiment}\ }\textbf {\bibinfo {volume} {2019}},\ \bibinfo {pages} {083103} (\bibinfo {year} {2019})}\BibitemShut {NoStop}%
\bibitem [{\citenamefont {Moudgalya}\ \emph {et~al.}(2018)\citenamefont {Moudgalya}, \citenamefont {Regnault},\ and\ \citenamefont {Bernevig}}]{AKLT2}%
  \BibitemOpen
  \bibfield  {author} {\bibinfo {author} {\bibfnamefont {S.}~\bibnamefont {Moudgalya}}, \bibinfo {author} {\bibfnamefont {N.}~\bibnamefont {Regnault}}, \ and\ \bibinfo {author} {\bibfnamefont {B.~A.}\ \bibnamefont {Bernevig}},\ }\bibfield  {title} {\textit {\bibinfo {title} {Entanglement of exact excited states of Affleck-Kennedy-Lieb-Tasaki models: Exact results, many-body scars, and violation of the strong eigenstate thermalization hypothesis},\ }}\href {\doibase 10.1103/PhysRevB.98.235156} {\bibfield  {journal} {\bibinfo  {journal} {Phys. Rev. B}\ }\textbf {\bibinfo {volume} {98}},\ \bibinfo {pages} {235156} (\bibinfo {year} {2018})}\BibitemShut {NoStop}%
\bibitem [{\citenamefont {Mark}\ \emph {et~al.}(2020)\citenamefont {Mark}, \citenamefont {Lin},\ and\ \citenamefont {Motrunich}}]{AKLT3}%
  \BibitemOpen
  \bibfield  {author} {\bibinfo {author} {\bibfnamefont {D.~K.}\ \bibnamefont {Mark}}, \bibinfo {author} {\bibfnamefont {C.-J.}\ \bibnamefont {Lin}}, \ and\ \bibinfo {author} {\bibfnamefont {O.~I.}\ \bibnamefont {Motrunich}},\ }\bibfield  {title} {\textit {\bibinfo {title} {Unified structure for exact towers of scar states in the Affleck-Kennedy-Lieb-Tasaki and other models},\ }}\href {\doibase 10.1103/PhysRevB.101.195131} {\bibfield  {journal} {\bibinfo  {journal} {Phys. Rev. B}\ }\textbf {\bibinfo {volume} {101}},\ \bibinfo {pages} {195131} (\bibinfo {year} {2020})}\BibitemShut {NoStop}%
\bibitem [{\citenamefont {Moudgalya}\ \emph {et~al.}(2020{\natexlab{c}})\citenamefont {Moudgalya}, \citenamefont {O'Brien}, \citenamefont {Bernevig}, \citenamefont {Fendley},\ and\ \citenamefont {Regnault}}]{AKLT4}%
  \BibitemOpen
  \bibfield  {author} {\bibinfo {author} {\bibfnamefont {S.}~\bibnamefont {Moudgalya}}, \bibinfo {author} {\bibfnamefont {E.}~\bibnamefont {O'Brien}}, \bibinfo {author} {\bibfnamefont {B.~A.}\ \bibnamefont {Bernevig}}, \bibinfo {author} {\bibfnamefont {P.}~\bibnamefont {Fendley}}, \ and\ \bibinfo {author} {\bibfnamefont {N.}~\bibnamefont {Regnault}},\ }\bibfield  {title} {\textit {\bibinfo {title} {Large classes of quantum scarred Hamiltonians from matrix product states},\ }}\href {\doibase 10.1103/PhysRevB.102.085120} {\bibfield  {journal} {\bibinfo  {journal} {Phys. Rev. B}\ }\textbf {\bibinfo {volume} {102}},\ \bibinfo {pages} {085120} (\bibinfo {year} {2020}{\natexlab{c}})}\BibitemShut {NoStop}%
\bibitem [{\citenamefont {Schecter}\ and\ \citenamefont {Iadecola}(2019)}]{XY1}%
  \BibitemOpen
  \bibfield  {author} {\bibinfo {author} {\bibfnamefont {M.}~\bibnamefont {Schecter}}\ and\ \bibinfo {author} {\bibfnamefont {T.}~\bibnamefont {Iadecola}},\ }\bibfield  {title} {\textit {\bibinfo {title} {Weak Ergodicity Breaking and Quantum Many-Body Scars in Spin-1 $XY$ Magnets},\ }}\href {\doibase 10.1103/PhysRevLett.123.147201} {\bibfield  {journal} {\bibinfo  {journal} {Phys. Rev. Lett.}\ }\textbf {\bibinfo {volume} {123}},\ \bibinfo {pages} {147201} (\bibinfo {year} {2019})}\BibitemShut {NoStop}%
\bibitem [{\citenamefont {Chattopadhyay}\ \emph {et~al.}(2020)\citenamefont {Chattopadhyay}, \citenamefont {Pichler}, \citenamefont {Lukin},\ and\ \citenamefont {Ho}}]{XY2}%
  \BibitemOpen
  \bibfield  {author} {\bibinfo {author} {\bibfnamefont {S.}~\bibnamefont {Chattopadhyay}}, \bibinfo {author} {\bibfnamefont {H.}~\bibnamefont {Pichler}}, \bibinfo {author} {\bibfnamefont {M.~D.}\ \bibnamefont {Lukin}}, \ and\ \bibinfo {author} {\bibfnamefont {W.~W.}\ \bibnamefont {Ho}},\ }\bibfield  {title} {\textit {\bibinfo {title} {Quantum many-body scars from virtual entangled pairs},\ }}\href {\doibase 10.1103/PhysRevB.101.174308} {\bibfield  {journal} {\bibinfo  {journal} {Phys. Rev. B}\ }\textbf {\bibinfo {volume} {101}},\ \bibinfo {pages} {174308} (\bibinfo {year} {2020})}\BibitemShut {NoStop}%
\bibitem [{\citenamefont {Lee}\ \emph {et~al.}(2020)\citenamefont {Lee}, \citenamefont {Melendrez}, \citenamefont {Pal},\ and\ \citenamefont {Changlani}}]{FrustratedSpin1}%
  \BibitemOpen
  \bibfield  {author} {\bibinfo {author} {\bibfnamefont {K.}~\bibnamefont {Lee}}, \bibinfo {author} {\bibfnamefont {R.}~\bibnamefont {Melendrez}}, \bibinfo {author} {\bibfnamefont {A.}~\bibnamefont {Pal}}, \ and\ \bibinfo {author} {\bibfnamefont {H.~J.}\ \bibnamefont {Changlani}},\ }\bibfield  {title} {\textit {\bibinfo {title} {Exact three-colored quantum scars from geometric frustration},\ }}\href {\doibase 10.1103/PhysRevB.101.241111} {\bibfield  {journal} {\bibinfo  {journal} {Phys. Rev. B}\ }\textbf {\bibinfo {volume} {101}},\ \bibinfo {pages} {241111} (\bibinfo {year} {2020})}\BibitemShut {NoStop}%
\bibitem [{\citenamefont {McClarty}\ \emph {et~al.}(2020)\citenamefont {McClarty}, \citenamefont {Haque}, \citenamefont {Sen},\ and\ \citenamefont {Richter}}]{FrustratedSpin2}%
  \BibitemOpen
  \bibfield  {author} {\bibinfo {author} {\bibfnamefont {P.~A.}\ \bibnamefont {McClarty}}, \bibinfo {author} {\bibfnamefont {M.}~\bibnamefont {Haque}}, \bibinfo {author} {\bibfnamefont {A.}~\bibnamefont {Sen}}, \ and\ \bibinfo {author} {\bibfnamefont {J.}~\bibnamefont {Richter}},\ }\bibfield  {title} {\textit {\bibinfo {title} {Disorder-free localization and many-body quantum scars from magnetic frustration},\ }}\href {\doibase 10.1103/PhysRevB.102.224303} {\bibfield  {journal} {\bibinfo  {journal} {Phys. Rev. B}\ }\textbf {\bibinfo {volume} {102}},\ \bibinfo {pages} {224303} (\bibinfo {year} {2020})}\BibitemShut {NoStop}%
\bibitem [{\citenamefont {Chandran}\ \emph {et~al.}(2023)\citenamefont {Chandran}, \citenamefont {Iadecola}, \citenamefont {Khemani},\ and\ \citenamefont {Moessner}}]{Review1}%
  \BibitemOpen
  \bibfield  {author} {\bibinfo {author} {\bibfnamefont {A.}~\bibnamefont {Chandran}}, \bibinfo {author} {\bibfnamefont {T.}~\bibnamefont {Iadecola}}, \bibinfo {author} {\bibfnamefont {V.}~\bibnamefont {Khemani}}, \ and\ \bibinfo {author} {\bibfnamefont {R.}~\bibnamefont {Moessner}},\ }\bibfield  {title} {\textit {\bibinfo {title} {Quantum Many-Body Scars: A Quasiparticle Perspective},\ }}\href {\doibase https://doi.org/10.1146/annurev-conmatphys-031620-101617} {\bibfield  {journal} {\bibinfo  {journal} {Annual Review of Condensed Matter Physics}\ }\textbf {\bibinfo {volume} {14}},\ \bibinfo {pages} {443} (\bibinfo {year} {2023})}\BibitemShut {NoStop}%
\bibitem [{\citenamefont {Moudgalya}\ \emph {et~al.}(2022)\citenamefont {Moudgalya}, \citenamefont {Bernevig},\ and\ \citenamefont {Regnault}}]{Review2}%
  \BibitemOpen
  \bibfield  {author} {\bibinfo {author} {\bibfnamefont {S.}~\bibnamefont {Moudgalya}}, \bibinfo {author} {\bibfnamefont {B.~A.}\ \bibnamefont {Bernevig}}, \ and\ \bibinfo {author} {\bibfnamefont {N.}~\bibnamefont {Regnault}},\ }\bibfield  {title} {\textit {\bibinfo {title} {Quantum many-body scars and Hilbert space fragmentation: a review of exact results},\ }}\href {\doibase 10.1088/1361-6633/ac73a0} {\bibfield  {journal} {\bibinfo  {journal} {Reports on Progress in Physics}\ }\textbf {\bibinfo {volume} {85}},\ \bibinfo {pages} {086501} (\bibinfo {year} {2022})}\BibitemShut {NoStop}%
\bibitem [{\citenamefont {Mondragon-Shem}\ \emph {et~al.}(2021)\citenamefont {Mondragon-Shem}, \citenamefont {Vavilov},\ and\ \citenamefont {Martin}}]{DisorderDetection_QSimulator_SPXP}%
  \BibitemOpen
  \bibfield  {author} {\bibinfo {author} {\bibfnamefont {I.}~\bibnamefont {Mondragon-Shem}}, \bibinfo {author} {\bibfnamefont {M.~G.}\ \bibnamefont {Vavilov}}, \ and\ \bibinfo {author} {\bibfnamefont {I.}~\bibnamefont {Martin}},\ }\bibfield  {title} {\textit {\bibinfo {title} {Fate of Quantum Many-Body Scars in the Presence of Disorder},\ }}\href {\doibase 10.1103/PRXQuantum.2.030349} {\bibfield  {journal} {\bibinfo  {journal} {PRX Quantum}\ }\textbf {\bibinfo {volume} {2}},\ \bibinfo {pages} {030349} (\bibinfo {year} {2021})}\BibitemShut {NoStop}%
\bibitem [{\citenamefont {Dooley}(2021)}]{QuantumSensing}%
  \BibitemOpen
  \bibfield  {author} {\bibinfo {author} {\bibfnamefont {S.}~\bibnamefont {Dooley}},\ }\bibfield  {title} {\textit {\bibinfo {title} {Robust Quantum Sensing in Strongly Interacting Systems with Many-Body Scars},\ }}\href {\doibase 10.1103/PRXQuantum.2.020330} {\bibfield  {journal} {\bibinfo  {journal} {PRX Quantum}\ }\textbf {\bibinfo {volume} {2}},\ \bibinfo {pages} {020330} (\bibinfo {year} {2021})}\BibitemShut {NoStop}%
\bibitem [{\citenamefont {Granovskii}\ and\ \citenamefont {Zhedanov}(1985{\natexlab{a}})}]{GranovskiiZhedanov1}%
  \BibitemOpen
  \bibfield  {author} {\bibinfo {author} {\bibfnamefont {Y.~I.}\ \bibnamefont {Granovskii}}\ and\ \bibinfo {author} {\bibfnamefont {A.}~\bibnamefont {Zhedanov}},\ }\bibfield  {title} {\textit {\bibinfo {title} {Periodic structures on a quantum spin chain},\ }}\href {http://jetp.ras.ru/cgi-bin/dn/e_062_06_1244.pdf} {\bibfield  {journal} {\bibinfo  {journal} {Zh. Eksp. Teor. Fiz}\ }\textbf {\bibinfo {volume} {89}},\ \bibinfo {pages} {2156} (\bibinfo {year} {1985}{\natexlab{a}})}\BibitemShut {NoStop}%
\bibitem [{\citenamefont {Granovskii}\ and\ \citenamefont {Zhedanov}(1985{\natexlab{b}})}]{GranovskiiZhedanov2}%
  \BibitemOpen
  \bibfield  {author} {\bibinfo {author} {\bibfnamefont {Y.~I.}\ \bibnamefont {Granovskii}}\ and\ \bibinfo {author} {\bibfnamefont {A.~S.}\ \bibnamefont {Zhedanov}},\ }\bibfield  {title} {\textit {\bibinfo {title} {Coherent structures in a Heisenberg anisotropic array},\ }}\href {http://jetpletters.ru/ps/0/article_22347.shtml} {\bibfield  {journal} {\bibinfo  {journal} {JETP Letters}\ }\textbf {\bibinfo {volume} {41}},\ \bibinfo {pages} {312} (\bibinfo {year} {1985}{\natexlab{b}})}\BibitemShut {NoStop}%
\bibitem [{\citenamefont {Bhowmick}\ \emph {et~al.}(2025)\citenamefont {Bhowmick}, \citenamefont {Bulchandani},\ and\ \citenamefont {Ho}}]{AssyemtricDecay}%
  \BibitemOpen
  \bibfield  {author} {\bibinfo {author} {\bibfnamefont {D.}~\bibnamefont {Bhowmick}}, \bibinfo {author} {\bibfnamefont {V.~B.}\ \bibnamefont {Bulchandani}}, \ and\ \bibinfo {author} {\bibfnamefont {W.~W.}\ \bibnamefont {Ho}},\ }\href@noop {} {\bibinfo {title} {Asymmetric decay of quantum many-body scars in XYZ quantum spin chains},\ } (\bibinfo {year} {2025}),\ \Eprint {http://arxiv.org/abs/2505.05435} {arXiv:2505.05435 [quant-ph]} \BibitemShut {NoStop}%
\bibitem [{\citenamefont {Zheng}\ \emph {et~al.}(2025)\citenamefont {Zheng}, \citenamefont {Liang}, \citenamefont {Chen},\ and\ \citenamefont {Zhang}}]{ExactSpinHelix}%
  \BibitemOpen
  \bibfield  {author} {\bibinfo {author} {\bibfnamefont {M.}~\bibnamefont {Zheng}}, \bibinfo {author} {\bibfnamefont {C.}~\bibnamefont {Liang}}, \bibinfo {author} {\bibfnamefont {S.}~\bibnamefont {Chen}}, \ and\ \bibinfo {author} {\bibfnamefont {X.}~\bibnamefont {Zhang}},\ }\bibfield  {title} {\textit {\bibinfo {title} {Exact spin helix eigenstates in the anisotropic spin-$s$ Heisenberg model with arbitrary dimensions},\ }}\href {\doibase 10.1103/6dbn-n6rv} {\bibfield  {journal} {\bibinfo  {journal} {Phys. Rev. B}\ }\textbf {\bibinfo {volume} {112}},\ \bibinfo {pages} {165102} (\bibinfo {year} {2025})}\BibitemShut {NoStop}%
\bibitem [{\citenamefont {Bhowmick}\ and\ \citenamefont {Ho}(2025)}]{GZ_ModernAlgebra}%
  \BibitemOpen
  \bibfield  {author} {\bibinfo {author} {\bibfnamefont {D.}~\bibnamefont {Bhowmick}}\ and\ \bibinfo {author} {\bibfnamefont {W.~W.}\ \bibnamefont {Ho}},\ }\href {https://arxiv.org/abs/2507.14895} {\bibinfo {title} {Granovskii-Zhedanov Scar of XYZ Spin-chain: Modern Algebraic Perspectives and Realization in Higher Dimensional Lattices},\ } (\bibinfo {year} {2025}),\ \Eprint {http://arxiv.org/abs/2507.14895} {arXiv:2507.14895 [quant-ph]} \BibitemShut {NoStop}%
\bibitem [{\citenamefont {Popkov}\ \emph {et~al.}(2021)\citenamefont {Popkov}, \citenamefont {Zhang},\ and\ \citenamefont {Kl\"umper}}]{BethePhantom1}%
  \BibitemOpen
  \bibfield  {author} {\bibinfo {author} {\bibfnamefont {V.}~\bibnamefont {Popkov}}, \bibinfo {author} {\bibfnamefont {X.}~\bibnamefont {Zhang}}, \ and\ \bibinfo {author} {\bibfnamefont {A.}~\bibnamefont {Kl\"umper}},\ }\bibfield  {title} {\textit {\bibinfo {title} {Phantom Bethe excitations and spin helix eigenstates in integrable periodic and open spin chains},\ }}\href {\doibase 10.1103/PhysRevB.104.L081410} {\bibfield  {journal} {\bibinfo  {journal} {Phys. Rev. B}\ }\textbf {\bibinfo {volume} {104}},\ \bibinfo {pages} {L081410} (\bibinfo {year} {2021})}\BibitemShut {NoStop}%
\bibitem [{\citenamefont {Jepsen}\ \emph {et~al.}(2022)\citenamefont {Jepsen}, \citenamefont {Lee}, \citenamefont {Lin}, \citenamefont {Dimitrova}, \citenamefont {Margalit}, \citenamefont {Ho},\ and\ \citenamefont {Ketterle}}]{BethePhantom2}%
  \BibitemOpen
  \bibfield  {author} {\bibinfo {author} {\bibfnamefont {P.~N.}\ \bibnamefont {Jepsen}}, \bibinfo {author} {\bibfnamefont {Y.~K.~E.}\ \bibnamefont {Lee}}, \bibinfo {author} {\bibfnamefont {H.}~\bibnamefont {Lin}}, \bibinfo {author} {\bibfnamefont {I.}~\bibnamefont {Dimitrova}}, \bibinfo {author} {\bibfnamefont {Y.}~\bibnamefont {Margalit}}, \bibinfo {author} {\bibfnamefont {W.~W.}\ \bibnamefont {Ho}}, \ and\ \bibinfo {author} {\bibfnamefont {W.}~\bibnamefont {Ketterle}},\ }\bibfield  {title} {\textit {\bibinfo {title} {Long-lived phantom helix states in Heisenberg quantum magnets},\ }}\href {\doibase 10.1038/s41567-022-01651-7} {\bibfield  {journal} {\bibinfo  {journal} {Nature Physics}\ }\textbf {\bibinfo {volume} {18}},\ \bibinfo {pages} {899–904} (\bibinfo {year} {2022})}\BibitemShut {NoStop}%
\bibitem [{\citenamefont {Hild}\ \emph {et~al.}(2014)\citenamefont {Hild}, \citenamefont {Fukuhara}, \citenamefont {Schau\ss{}}, \citenamefont {Zeiher}, \citenamefont {Knap}, \citenamefont {Demler}, \citenamefont {Bloch},\ and\ \citenamefont {Gross}}]{BethePhantom3}%
  \BibitemOpen
  \bibfield  {author} {\bibinfo {author} {\bibfnamefont {S.}~\bibnamefont {Hild}}, \bibinfo {author} {\bibfnamefont {T.}~\bibnamefont {Fukuhara}}, \bibinfo {author} {\bibfnamefont {P.}~\bibnamefont {Schau\ss{}}}, \bibinfo {author} {\bibfnamefont {J.}~\bibnamefont {Zeiher}}, \bibinfo {author} {\bibfnamefont {M.}~\bibnamefont {Knap}}, \bibinfo {author} {\bibfnamefont {E.}~\bibnamefont {Demler}}, \bibinfo {author} {\bibfnamefont {I.}~\bibnamefont {Bloch}}, \ and\ \bibinfo {author} {\bibfnamefont {C.}~\bibnamefont {Gross}},\ }\bibfield  {title} {\textit {\bibinfo {title} {Far-from-Equilibrium Spin Transport in Heisenberg Quantum Magnets},\ }}\href {\doibase 10.1103/PhysRevLett.113.147205} {\bibfield  {journal} {\bibinfo  {journal} {Phys. Rev. Lett.}\ }\textbf {\bibinfo {volume} {113}},\ \bibinfo {pages} {147205} (\bibinfo {year} {2014})}\BibitemShut {NoStop}%
\bibitem [{\citenamefont {Zhang}\ and\ \citenamefont {Song}(2024)}]{BethePhantom5}%
  \BibitemOpen
  \bibfield  {author} {\bibinfo {author} {\bibfnamefont {G.}~\bibnamefont {Zhang}}\ and\ \bibinfo {author} {\bibfnamefont {Z.}~\bibnamefont {Song}},\ }\bibfield  {title} {\textit {\bibinfo {title} {Stable dynamic helix state in the nonintegrable XXZ Heisenberg model},\ }}\href {\doibase 10.1088/1402-4896/ad5886} {\bibfield  {journal} {\bibinfo  {journal} {Physica Scripta}\ }\textbf {\bibinfo {volume} {99}},\ \bibinfo {pages} {075119} (\bibinfo {year} {2024})}\BibitemShut {NoStop}%
\bibitem [{\citenamefont {Lin}\ \emph {et~al.}(2020)\citenamefont {Lin}, \citenamefont {Chandran},\ and\ \citenamefont {Motrunich}}]{SlowThermalization}%
  \BibitemOpen
  \bibfield  {author} {\bibinfo {author} {\bibfnamefont {C.-J.}\ \bibnamefont {Lin}}, \bibinfo {author} {\bibfnamefont {A.}~\bibnamefont {Chandran}}, \ and\ \bibinfo {author} {\bibfnamefont {O.~I.}\ \bibnamefont {Motrunich}},\ }\bibfield  {title} {\textit {\bibinfo {title} {Slow thermalization of exact quantum many-body scar states under perturbations},\ }}\href {\doibase 10.1103/PhysRevResearch.2.033044} {\bibfield  {journal} {\bibinfo  {journal} {Phys. Rev. Res.}\ }\textbf {\bibinfo {volume} {2}},\ \bibinfo {pages} {033044} (\bibinfo {year} {2020})}\BibitemShut {NoStop}%
\bibitem [{\citenamefont {Bluvstein}\ \emph {et~al.}(2021)\citenamefont {Bluvstein}, \citenamefont {Omran}, \citenamefont {Levine}, \citenamefont {Keesling}, \citenamefont {Semeghini}, \citenamefont {Ebadi}, \citenamefont {Wang}, \citenamefont {Michailidis}, \citenamefont {Maskara}, \citenamefont {Ho}, \citenamefont {Choi}, \citenamefont {Serbyn}, \citenamefont {Greiner}, \citenamefont {Vuletić},\ and\ \citenamefont {Lukin}}]{DrivingProtocol1}%
  \BibitemOpen
  \bibfield  {author} {\bibinfo {author} {\bibfnamefont {D.}~\bibnamefont {Bluvstein}}, \bibinfo {author} {\bibfnamefont {A.}~\bibnamefont {Omran}}, \bibinfo {author} {\bibfnamefont {H.}~\bibnamefont {Levine}}, \bibinfo {author} {\bibfnamefont {A.}~\bibnamefont {Keesling}}, \bibinfo {author} {\bibfnamefont {G.}~\bibnamefont {Semeghini}}, \bibinfo {author} {\bibfnamefont {S.}~\bibnamefont {Ebadi}}, \bibinfo {author} {\bibfnamefont {T.~T.}\ \bibnamefont {Wang}}, \bibinfo {author} {\bibfnamefont {A.~A.}\ \bibnamefont {Michailidis}}, \bibinfo {author} {\bibfnamefont {N.}~\bibnamefont {Maskara}}, \bibinfo {author} {\bibfnamefont {W.~W.}\ \bibnamefont {Ho}}, \bibinfo {author} {\bibfnamefont {S.}~\bibnamefont {Choi}}, \bibinfo {author} {\bibfnamefont {M.}~\bibnamefont {Serbyn}}, \bibinfo {author} {\bibfnamefont {M.}~\bibnamefont {Greiner}}, \bibinfo {author} {\bibfnamefont {V.}~\bibnamefont {Vuletić}}, \ and\ \bibinfo {author} {\bibfnamefont {M.~D.}\ \bibnamefont {Lukin}},\ }\bibfield  {title} {\textit {\bibinfo
  {title} {Controlling quantum many-body dynamics in driven Rydberg atom arrays},\ }}\href {\doibase 10.1126/science.abg2530} {\bibfield  {journal} {\bibinfo  {journal} {Science}\ }\textbf {\bibinfo {volume} {371}},\ \bibinfo {pages} {1355} (\bibinfo {year} {2021})}\BibitemShut {NoStop}%
\bibitem [{\citenamefont {Su}\ \emph {et~al.}(2023)\citenamefont {Su}, \citenamefont {Sun}, \citenamefont {Hudomal}, \citenamefont {Desaules}, \citenamefont {Zhou}, \citenamefont {Yang}, \citenamefont {Halimeh}, \citenamefont {Yuan}, \citenamefont {Papi\ifmmode~\acute{c}\else \'{c}\fi{}},\ and\ \citenamefont {Pan}}]{DrivingProtocol2}%
  \BibitemOpen
  \bibfield  {author} {\bibinfo {author} {\bibfnamefont {G.-X.}\ \bibnamefont {Su}}, \bibinfo {author} {\bibfnamefont {H.}~\bibnamefont {Sun}}, \bibinfo {author} {\bibfnamefont {A.}~\bibnamefont {Hudomal}}, \bibinfo {author} {\bibfnamefont {J.-Y.}\ \bibnamefont {Desaules}}, \bibinfo {author} {\bibfnamefont {Z.-Y.}\ \bibnamefont {Zhou}}, \bibinfo {author} {\bibfnamefont {B.}~\bibnamefont {Yang}}, \bibinfo {author} {\bibfnamefont {J.~C.}\ \bibnamefont {Halimeh}}, \bibinfo {author} {\bibfnamefont {Z.-S.}\ \bibnamefont {Yuan}}, \bibinfo {author} {\bibfnamefont {Z.}~\bibnamefont {Papi\ifmmode~\acute{c}\else \'{c}\fi{}}}, \ and\ \bibinfo {author} {\bibfnamefont {J.-W.}\ \bibnamefont {Pan}},\ }\bibfield  {title} {\textit {\bibinfo {title} {Observation of many-body scarring in a Bose-Hubbard quantum simulator},\ }}\href {\doibase 10.1103/PhysRevResearch.5.023010} {\bibfield  {journal} {\bibinfo  {journal} {Phys. Rev. Res.}\ }\textbf {\bibinfo {volume} {5}},\ \bibinfo {pages} {023010} (\bibinfo {year}
  {2023})}\BibitemShut {NoStop}%
\bibitem [{\citenamefont {Maskara}\ \emph {et~al.}(2021)\citenamefont {Maskara}, \citenamefont {Michailidis}, \citenamefont {Ho}, \citenamefont {Bluvstein}, \citenamefont {Choi}, \citenamefont {Lukin},\ and\ \citenamefont {Serbyn}}]{DrivingProtocol3}%
  \BibitemOpen
  \bibfield  {author} {\bibinfo {author} {\bibfnamefont {N.}~\bibnamefont {Maskara}}, \bibinfo {author} {\bibfnamefont {A.~A.}\ \bibnamefont {Michailidis}}, \bibinfo {author} {\bibfnamefont {W.~W.}\ \bibnamefont {Ho}}, \bibinfo {author} {\bibfnamefont {D.}~\bibnamefont {Bluvstein}}, \bibinfo {author} {\bibfnamefont {S.}~\bibnamefont {Choi}}, \bibinfo {author} {\bibfnamefont {M.~D.}\ \bibnamefont {Lukin}}, \ and\ \bibinfo {author} {\bibfnamefont {M.}~\bibnamefont {Serbyn}},\ }\bibfield  {title} {\textit {\bibinfo {title} {Discrete Time-Crystalline Order Enabled by Quantum Many-Body Scars: Entanglement Steering via Periodic Driving},\ }}\href {\doibase 10.1103/PhysRevLett.127.090602} {\bibfield  {journal} {\bibinfo  {journal} {Phys. Rev. Lett.}\ }\textbf {\bibinfo {volume} {127}},\ \bibinfo {pages} {090602} (\bibinfo {year} {2021})}\BibitemShut {NoStop}%
\bibitem [{\citenamefont {Hudomal}\ \emph {et~al.}(2022)\citenamefont {Hudomal}, \citenamefont {Desaules}, \citenamefont {Mukherjee}, \citenamefont {Su}, \citenamefont {Halimeh},\ and\ \citenamefont {Papi\ifmmode~\acute{c}\else \'{c}\fi{}}}]{DrivingProtocol4}%
  \BibitemOpen
  \bibfield  {author} {\bibinfo {author} {\bibfnamefont {A.}~\bibnamefont {Hudomal}}, \bibinfo {author} {\bibfnamefont {J.-Y.}\ \bibnamefont {Desaules}}, \bibinfo {author} {\bibfnamefont {B.}~\bibnamefont {Mukherjee}}, \bibinfo {author} {\bibfnamefont {G.-X.}\ \bibnamefont {Su}}, \bibinfo {author} {\bibfnamefont {J.~C.}\ \bibnamefont {Halimeh}}, \ and\ \bibinfo {author} {\bibfnamefont {Z.}~\bibnamefont {Papi\ifmmode~\acute{c}\else \'{c}\fi{}}},\ }\bibfield  {title} {\textit {\bibinfo {title} {Driving quantum many-body scars in the PXP model},\ }}\href {\doibase 10.1103/PhysRevB.106.104302} {\bibfield  {journal} {\bibinfo  {journal} {Phys. Rev. B}\ }\textbf {\bibinfo {volume} {106}},\ \bibinfo {pages} {104302} (\bibinfo {year} {2022})}\BibitemShut {NoStop}%
\bibitem [{\citenamefont {Halimeh}\ \emph {et~al.}(2023)\citenamefont {Halimeh}, \citenamefont {Barbiero}, \citenamefont {Hauke}, \citenamefont {Grusdt},\ and\ \citenamefont {Bohrdt}}]{DrivingProtocol5}%
  \BibitemOpen
  \bibfield  {author} {\bibinfo {author} {\bibfnamefont {J.~C.}\ \bibnamefont {Halimeh}}, \bibinfo {author} {\bibfnamefont {L.}~\bibnamefont {Barbiero}}, \bibinfo {author} {\bibfnamefont {P.}~\bibnamefont {Hauke}}, \bibinfo {author} {\bibfnamefont {F.}~\bibnamefont {Grusdt}}, \ and\ \bibinfo {author} {\bibfnamefont {A.}~\bibnamefont {Bohrdt}},\ }\bibfield  {title} {\textit {\bibinfo {title} {Robust quantum many-body scars in lattice gauge theories},\ }}\href {\doibase 10.22331/q-2023-05-15-1004} {\bibfield  {journal} {\bibinfo  {journal} {{Quantum}}\ }\textbf {\bibinfo {volume} {7}},\ \bibinfo {pages} {1004} (\bibinfo {year} {2023})}\BibitemShut {NoStop}%
\bibitem [{\citenamefont {Shen}\ \emph {et~al.}(2024)\citenamefont {Shen}, \citenamefont {Qin}, \citenamefont {Desaules}, \citenamefont {Papi\ifmmode~\acute{c}\else \'{c}\fi{}},\ and\ \citenamefont {Lee}}]{Stabilize2}%
  \BibitemOpen
  \bibfield  {author} {\bibinfo {author} {\bibfnamefont {R.}~\bibnamefont {Shen}}, \bibinfo {author} {\bibfnamefont {F.}~\bibnamefont {Qin}}, \bibinfo {author} {\bibfnamefont {J.-Y.}\ \bibnamefont {Desaules}}, \bibinfo {author} {\bibfnamefont {Z.}~\bibnamefont {Papi\ifmmode~\acute{c}\else \'{c}\fi{}}}, \ and\ \bibinfo {author} {\bibfnamefont {C.~H.}\ \bibnamefont {Lee}},\ }\bibfield  {title} {\textit {\bibinfo {title} {Enhanced Many-Body Quantum Scars from the Non-Hermitian Fock Skin Effect},\ }}\href {\doibase 10.1103/PhysRevLett.133.216601} {\bibfield  {journal} {\bibinfo  {journal} {Phys. Rev. Lett.}\ }\textbf {\bibinfo {volume} {133}},\ \bibinfo {pages} {216601} (\bibinfo {year} {2024})}\BibitemShut {NoStop}%
\bibitem [{\citenamefont {Jiang}\ \emph {et~al.}(2026)\citenamefont {Jiang}, \citenamefont {Xu}, \citenamefont {Yang}, \citenamefont {Hou}, \citenamefont {Wang},\ and\ \citenamefont {Pan}}]{Stabilize3}%
  \BibitemOpen
  \bibfield  {author} {\bibinfo {author} {\bibfnamefont {X.-P.}\ \bibnamefont {Jiang}}, \bibinfo {author} {\bibfnamefont {M.}~\bibnamefont {Xu}}, \bibinfo {author} {\bibfnamefont {X.}~\bibnamefont {Yang}}, \bibinfo {author} {\bibfnamefont {H.}~\bibnamefont {Hou}}, \bibinfo {author} {\bibfnamefont {Y.}~\bibnamefont {Wang}}, \ and\ \bibinfo {author} {\bibfnamefont {L.}~\bibnamefont {Pan}},\ }\bibfield  {title} {\textit {\bibinfo {title} {{Robustness of quantum many-body scars in the presence of a Markovian bath}},\ }}\href {\doibase 10.1038/s42005-025-02446-x} {\bibfield  {journal} {\bibinfo  {journal} {Commun. Phys.}\ }\textbf {\bibinfo {volume} {9}},\ \bibinfo {pages} {14} (\bibinfo {year} {2026})}\BibitemShut {NoStop}%
\bibitem [{\citenamefont {Ma}\ \emph {et~al.}(2022)\citenamefont {Ma}, \citenamefont {Zhang},\ and\ \citenamefont {Song}}]{HelixDynamics1}%
  \BibitemOpen
  \bibfield  {author} {\bibinfo {author} {\bibfnamefont {E.~S.}\ \bibnamefont {Ma}}, \bibinfo {author} {\bibfnamefont {K.~L.}\ \bibnamefont {Zhang}}, \ and\ \bibinfo {author} {\bibfnamefont {Z.}~\bibnamefont {Song}},\ }\bibfield  {title} {\textit {\bibinfo {title} {Steady helix states in a resonant XXZ Heisenberg model with Dzyaloshinskii-Moriya interaction},\ }}\href {\doibase 10.1103/PhysRevB.106.245122} {\bibfield  {journal} {\bibinfo  {journal} {Phys. Rev. B}\ }\textbf {\bibinfo {volume} {106}},\ \bibinfo {pages} {245122} (\bibinfo {year} {2022})}\BibitemShut {NoStop}%
\bibitem [{\citenamefont {Shi}\ and\ \citenamefont {Song}(2023)}]{SGA2}%
  \BibitemOpen
  \bibfield  {author} {\bibinfo {author} {\bibfnamefont {Y.~B.}\ \bibnamefont {Shi}}\ and\ \bibinfo {author} {\bibfnamefont {Z.}~\bibnamefont {Song}},\ }\bibfield  {title} {\textit {\bibinfo {title} {Robust unidirectional phantom helix states in the XXZ Heisenberg model with Dzyaloshinskii-Moriya interaction},\ }}\href {\doibase 10.1103/PhysRevB.108.085108} {\bibfield  {journal} {\bibinfo  {journal} {Phys. Rev. B}\ }\textbf {\bibinfo {volume} {108}},\ \bibinfo {pages} {085108} (\bibinfo {year} {2023})}\BibitemShut {NoStop}%
\bibitem [{\citenamefont {Popkov}\ and\ \citenamefont {Presilla}(2016)}]{VladislavPopkov1}%
  \BibitemOpen
  \bibfield  {author} {\bibinfo {author} {\bibfnamefont {V.}~\bibnamefont {Popkov}}\ and\ \bibinfo {author} {\bibfnamefont {C.}~\bibnamefont {Presilla}},\ }\bibfield  {title} {\textit {\bibinfo {title} {Obtaining pure steady states in nonequilibrium quantum systems with strong dissipative couplings},\ }}\href {\doibase 10.1103/PhysRevA.93.022111} {\bibfield  {journal} {\bibinfo  {journal} {Phys. Rev. A}\ }\textbf {\bibinfo {volume} {93}},\ \bibinfo {pages} {022111} (\bibinfo {year} {2016})}\BibitemShut {NoStop}%
\bibitem [{\citenamefont {Popkov}\ \emph {et~al.}(2017)\citenamefont {Popkov}, \citenamefont {Schmidt},\ and\ \citenamefont {Presilla}}]{VladislavPopkov2}%
  \BibitemOpen
  \bibfield  {author} {\bibinfo {author} {\bibfnamefont {V.}~\bibnamefont {Popkov}}, \bibinfo {author} {\bibfnamefont {J.}~\bibnamefont {Schmidt}}, \ and\ \bibinfo {author} {\bibfnamefont {C.}~\bibnamefont {Presilla}},\ }\bibfield  {title} {\textit {\bibinfo {title} {Spin-helix states in theXXZspin chain with strong boundary dissipation},\ }}\href {\doibase 10.1088/1751-8121/aa86cb} {\bibfield  {journal} {\bibinfo  {journal} {Journal of Physics A: Mathematical and Theoretical}\ }\textbf {\bibinfo {volume} {50}},\ \bibinfo {pages} {435302} (\bibinfo {year} {2017})}\BibitemShut {NoStop}%
\bibitem [{\citenamefont {Gerken}\ \emph {et~al.}(2025)\citenamefont {Gerken}, \citenamefont {Runkel}, \citenamefont {Schweigert},\ and\ \citenamefont {Posske}}]{FelixGraphicalConstruction}%
  \BibitemOpen
  \bibfield  {author} {\bibinfo {author} {\bibfnamefont {F.}~\bibnamefont {Gerken}}, \bibinfo {author} {\bibfnamefont {I.}~\bibnamefont {Runkel}}, \bibinfo {author} {\bibfnamefont {C.}~\bibnamefont {Schweigert}}, \ and\ \bibinfo {author} {\bibfnamefont {T.}~\bibnamefont {Posske}},\ }\bibfield  {title} {\textit {\bibinfo {title} {All product eigenstates in Heisenberg models from a graphical construction},\ }}\href {\doibase 10.1103/PhysRevResearch.7.L012008} {\bibfield  {journal} {\bibinfo  {journal} {Phys. Rev. Res.}\ }\textbf {\bibinfo {volume} {7}},\ \bibinfo {pages} {L012008} (\bibinfo {year} {2025})}\BibitemShut {NoStop}%
\bibitem [{\citenamefont {Hauschild}\ \emph {et~al.}(2024)\citenamefont {Hauschild}, \citenamefont {Unfried}, \citenamefont {Anand}, \citenamefont {Andrews}, \citenamefont {Bintz}, \citenamefont {Borla}, \citenamefont {Divic}, \citenamefont {Drescher}, \citenamefont {Geiger}, \citenamefont {Hefel}, \citenamefont {Hémery}, \citenamefont {Kadow}, \citenamefont {Kemp}, \citenamefont {Kirchner}, \citenamefont {Liu}, \citenamefont {Möller}, \citenamefont {Parker}, \citenamefont {Rader}, \citenamefont {Romen}, \citenamefont {Scalet}, \citenamefont {Schoonderwoerd}, \citenamefont {Schulz}, \citenamefont {Soejima}, \citenamefont {Thoma}, \citenamefont {Wu}, \citenamefont {Zechmann}, \citenamefont {Zweng}, \citenamefont {Mong}, \citenamefont {Zaletel},\ and\ \citenamefont {Pollmann}}]{tenpy2024}%
  \BibitemOpen
  \bibfield  {author} {\bibinfo {author} {\bibfnamefont {J.}~\bibnamefont {Hauschild}}, \bibinfo {author} {\bibfnamefont {J.}~\bibnamefont {Unfried}}, \bibinfo {author} {\bibfnamefont {S.}~\bibnamefont {Anand}}, \bibinfo {author} {\bibfnamefont {B.}~\bibnamefont {Andrews}}, \bibinfo {author} {\bibfnamefont {M.}~\bibnamefont {Bintz}}, \bibinfo {author} {\bibfnamefont {U.}~\bibnamefont {Borla}}, \bibinfo {author} {\bibfnamefont {S.}~\bibnamefont {Divic}}, \bibinfo {author} {\bibfnamefont {M.}~\bibnamefont {Drescher}}, \bibinfo {author} {\bibfnamefont {J.}~\bibnamefont {Geiger}}, \bibinfo {author} {\bibfnamefont {M.}~\bibnamefont {Hefel}}, \bibinfo {author} {\bibfnamefont {K.}~\bibnamefont {Hémery}}, \bibinfo {author} {\bibfnamefont {W.}~\bibnamefont {Kadow}}, \bibinfo {author} {\bibfnamefont {J.}~\bibnamefont {Kemp}}, \bibinfo {author} {\bibfnamefont {N.}~\bibnamefont {Kirchner}}, \bibinfo {author} {\bibfnamefont {V.~S.}\ \bibnamefont {Liu}}, \bibinfo {author} {\bibfnamefont {G.}~\bibnamefont {Möller}},
  \bibinfo {author} {\bibfnamefont {D.}~\bibnamefont {Parker}}, \bibinfo {author} {\bibfnamefont {M.}~\bibnamefont {Rader}}, \bibinfo {author} {\bibfnamefont {A.}~\bibnamefont {Romen}}, \bibinfo {author} {\bibfnamefont {S.}~\bibnamefont {Scalet}}, \bibinfo {author} {\bibfnamefont {L.}~\bibnamefont {Schoonderwoerd}}, \bibinfo {author} {\bibfnamefont {M.}~\bibnamefont {Schulz}}, \bibinfo {author} {\bibfnamefont {T.}~\bibnamefont {Soejima}}, \bibinfo {author} {\bibfnamefont {P.}~\bibnamefont {Thoma}}, \bibinfo {author} {\bibfnamefont {Y.}~\bibnamefont {Wu}}, \bibinfo {author} {\bibfnamefont {P.}~\bibnamefont {Zechmann}}, \bibinfo {author} {\bibfnamefont {L.}~\bibnamefont {Zweng}}, \bibinfo {author} {\bibfnamefont {R.~S.~K.}\ \bibnamefont {Mong}}, \bibinfo {author} {\bibfnamefont {M.~P.}\ \bibnamefont {Zaletel}}, \ and\ \bibinfo {author} {\bibfnamefont {F.}~\bibnamefont {Pollmann}},\ }\bibfield  {title} {\textit {\bibinfo {title} {{Tensor network Python (TeNPy) version 1}},\ }}\href {\doibase
  10.21468/SciPostPhysCodeb.41} {\bibfield  {journal} {\bibinfo  {journal} {SciPost Phys. Codebases}\ ,\ \bibinfo {pages} {41}} (\bibinfo {year} {2024})}\BibitemShut {NoStop}%
\bibitem [{\citenamefont {Schiulaz}\ \emph {et~al.}(2019)\citenamefont {Schiulaz}, \citenamefont {Torres-Herrera},\ and\ \citenamefont {Santos}}]{ThoulessTime}%
  \BibitemOpen
  \bibfield  {author} {\bibinfo {author} {\bibfnamefont {M.}~\bibnamefont {Schiulaz}}, \bibinfo {author} {\bibfnamefont {E.~J.}\ \bibnamefont {Torres-Herrera}}, \ and\ \bibinfo {author} {\bibfnamefont {L.~F.}\ \bibnamefont {Santos}},\ }\bibfield  {title} {\textit {\bibinfo {title} {Thouless and relaxation time scales in many-body quantum systems},\ }}\href {\doibase 10.1103/PhysRevB.99.174313} {\bibfield  {journal} {\bibinfo  {journal} {Phys. Rev. B}\ }\textbf {\bibinfo {volume} {99}},\ \bibinfo {pages} {174313} (\bibinfo {year} {2019})}\BibitemShut {NoStop}%
\bibitem [{\citenamefont {Lambert}\ \emph {et~al.}(2026)\citenamefont {Lambert}, \citenamefont {Giguère}, \citenamefont {Menczel}, \citenamefont {Li}, \citenamefont {Hopf}, \citenamefont {Suárez}, \citenamefont {Gali}, \citenamefont {Lishman}, \citenamefont {Gadhvi}, \citenamefont {Agarwal}, \citenamefont {Galicia}, \citenamefont {Shammah}, \citenamefont {Nation}, \citenamefont {Johansson}, \citenamefont {Ahmed}, \citenamefont {Cross}, \citenamefont {Pitchford},\ and\ \citenamefont {Nori}}]{QuTip}%
  \BibitemOpen
  \bibfield  {author} {\bibinfo {author} {\bibfnamefont {N.}~\bibnamefont {Lambert}}, \bibinfo {author} {\bibfnamefont {E.}~\bibnamefont {Giguère}}, \bibinfo {author} {\bibfnamefont {P.}~\bibnamefont {Menczel}}, \bibinfo {author} {\bibfnamefont {B.}~\bibnamefont {Li}}, \bibinfo {author} {\bibfnamefont {P.}~\bibnamefont {Hopf}}, \bibinfo {author} {\bibfnamefont {G.}~\bibnamefont {Suárez}}, \bibinfo {author} {\bibfnamefont {M.}~\bibnamefont {Gali}}, \bibinfo {author} {\bibfnamefont {J.}~\bibnamefont {Lishman}}, \bibinfo {author} {\bibfnamefont {R.}~\bibnamefont {Gadhvi}}, \bibinfo {author} {\bibfnamefont {R.}~\bibnamefont {Agarwal}}, \bibinfo {author} {\bibfnamefont {A.}~\bibnamefont {Galicia}}, \bibinfo {author} {\bibfnamefont {N.}~\bibnamefont {Shammah}}, \bibinfo {author} {\bibfnamefont {P.}~\bibnamefont {Nation}}, \bibinfo {author} {\bibfnamefont {J.}~\bibnamefont {Johansson}}, \bibinfo {author} {\bibfnamefont {S.}~\bibnamefont {Ahmed}}, \bibinfo {author} {\bibfnamefont {S.}~\bibnamefont {Cross}},
  \bibinfo {author} {\bibfnamefont {A.}~\bibnamefont {Pitchford}}, \ and\ \bibinfo {author} {\bibfnamefont {F.}~\bibnamefont {Nori}},\ }\bibfield  {title} {\textit {\bibinfo {title} {QuTiP 5: The Quantum Toolbox in Python},\ }}\href {\doibase https://doi.org/10.1016/j.physrep.2025.10.001} {\bibfield  {journal} {\bibinfo  {journal} {Physics Reports}\ }\textbf {\bibinfo {volume} {1153}},\ \bibinfo {pages} {1} (\bibinfo {year} {2026})},\ \bibinfo {note} {quTiP 5: The Quantum Toolbox in Python}\BibitemShut {NoStop}%
\bibitem [{\citenamefont {Bienfait}\ \emph {et~al.}(2016)\citenamefont {Bienfait}, \citenamefont {Pla}, \citenamefont {Kubo}, \citenamefont {Zhou}, \citenamefont {Stern}, \citenamefont {Lo}, \citenamefont {Weis}, \citenamefont {Schenkel}, \citenamefont {Vion}, \citenamefont {Esteve}, \citenamefont {Morton},\ and\ \citenamefont {Bertet}}]{PurcellEffect1}%
  \BibitemOpen
  \bibfield  {author} {\bibinfo {author} {\bibfnamefont {A.}~\bibnamefont {Bienfait}}, \bibinfo {author} {\bibfnamefont {J.~J.}\ \bibnamefont {Pla}}, \bibinfo {author} {\bibfnamefont {Y.}~\bibnamefont {Kubo}}, \bibinfo {author} {\bibfnamefont {X.}~\bibnamefont {Zhou}}, \bibinfo {author} {\bibfnamefont {M.}~\bibnamefont {Stern}}, \bibinfo {author} {\bibfnamefont {C.~C.}\ \bibnamefont {Lo}}, \bibinfo {author} {\bibfnamefont {C.~D.}\ \bibnamefont {Weis}}, \bibinfo {author} {\bibfnamefont {T.}~\bibnamefont {Schenkel}}, \bibinfo {author} {\bibfnamefont {D.}~\bibnamefont {Vion}}, \bibinfo {author} {\bibfnamefont {D.}~\bibnamefont {Esteve}}, \bibinfo {author} {\bibfnamefont {J.~J.~L.}\ \bibnamefont {Morton}}, \ and\ \bibinfo {author} {\bibfnamefont {P.}~\bibnamefont {Bertet}},\ }\bibfield  {title} {\textit {\bibinfo {title} {Controlling spin relaxation with a cavity},\ }}\href {\doibase 10.1038/nature16944} {\bibfield  {journal} {\bibinfo  {journal} {Nature}\ }\textbf {\bibinfo {volume} {531}},\ \bibinfo {pages}
  {74–77} (\bibinfo {year} {2016})}\BibitemShut {NoStop}%
\bibitem [{\citenamefont {Metzger}\ \emph {et~al.}(2019)\citenamefont {Metzger}, \citenamefont {Muller}, \citenamefont {Nishida}, \citenamefont {Pollard}, \citenamefont {Hentschel},\ and\ \citenamefont {Raschke}}]{PurcellEffect2}%
  \BibitemOpen
  \bibfield  {author} {\bibinfo {author} {\bibfnamefont {B.}~\bibnamefont {Metzger}}, \bibinfo {author} {\bibfnamefont {E.}~\bibnamefont {Muller}}, \bibinfo {author} {\bibfnamefont {J.}~\bibnamefont {Nishida}}, \bibinfo {author} {\bibfnamefont {B.}~\bibnamefont {Pollard}}, \bibinfo {author} {\bibfnamefont {M.}~\bibnamefont {Hentschel}}, \ and\ \bibinfo {author} {\bibfnamefont {M.~B.}\ \bibnamefont {Raschke}},\ }\bibfield  {title} {\textit {\bibinfo {title} {Purcell-Enhanced Spontaneous Emission of Molecular Vibrations},\ }}\href {\doibase 10.1103/PhysRevLett.123.153001} {\bibfield  {journal} {\bibinfo  {journal} {Phys. Rev. Lett.}\ }\textbf {\bibinfo {volume} {123}},\ \bibinfo {pages} {153001} (\bibinfo {year} {2019})}\BibitemShut {NoStop}%
\bibitem [{\citenamefont {Wang}\ \emph {et~al.}(2025)\citenamefont {Wang}, \citenamefont {Zhou}, \citenamefont {Shen}, \citenamefont {Huang}, \citenamefont {He}, \citenamefont {Huang}, \citenamefont {Liu}, \citenamefont {Li},\ and\ \citenamefont {Guo}}]{PurcellEffect3}%
  \BibitemOpen
  \bibfield  {author} {\bibinfo {author} {\bibfnamefont {J.}~\bibnamefont {Wang}}, \bibinfo {author} {\bibfnamefont {X.-L.}\ \bibnamefont {Zhou}}, \bibinfo {author} {\bibfnamefont {Z.-M.}\ \bibnamefont {Shen}}, \bibinfo {author} {\bibfnamefont {D.-Y.}\ \bibnamefont {Huang}}, \bibinfo {author} {\bibfnamefont {S.-J.}\ \bibnamefont {He}}, \bibinfo {author} {\bibfnamefont {Q.-Y.}\ \bibnamefont {Huang}}, \bibinfo {author} {\bibfnamefont {Y.-J.}\ \bibnamefont {Liu}}, \bibinfo {author} {\bibfnamefont {C.-F.}\ \bibnamefont {Li}}, \ and\ \bibinfo {author} {\bibfnamefont {G.-C.}\ \bibnamefont {Guo}},\ }\bibfield  {title} {\textit {\bibinfo {title} {Purcell-Enhanced Generation of Photonic Bell States via the Inelastic Scattering off Single Atoms},\ }}\href {\doibase 10.1103/PhysRevLett.134.053401} {\bibfield  {journal} {\bibinfo  {journal} {Phys. Rev. Lett.}\ }\textbf {\bibinfo {volume} {134}},\ \bibinfo {pages} {053401} (\bibinfo {year} {2025})}\BibitemShut {NoStop}%
\bibitem [{\citenamefont {Nagorny}\ \emph {et~al.}(2003)\citenamefont {Nagorny}, \citenamefont {Els\"asser}, \citenamefont {Richter}, \citenamefont {Hemmerich}, \citenamefont {Kruse}, \citenamefont {Zimmermann},\ and\ \citenamefont {Courteille}}]{RingOpticalLattice}%
  \BibitemOpen
  \bibfield  {author} {\bibinfo {author} {\bibfnamefont {B.}~\bibnamefont {Nagorny}}, \bibinfo {author} {\bibfnamefont {T.}~\bibnamefont {Els\"asser}}, \bibinfo {author} {\bibfnamefont {H.}~\bibnamefont {Richter}}, \bibinfo {author} {\bibfnamefont {A.}~\bibnamefont {Hemmerich}}, \bibinfo {author} {\bibfnamefont {D.}~\bibnamefont {Kruse}}, \bibinfo {author} {\bibfnamefont {C.}~\bibnamefont {Zimmermann}}, \ and\ \bibinfo {author} {\bibfnamefont {P.}~\bibnamefont {Courteille}},\ }\bibfield  {title} {\textit {\bibinfo {title} {Optical lattice in a high-finesse ring resonator},\ }}\href {\doibase 10.1103/PhysRevA.67.031401} {\bibfield  {journal} {\bibinfo  {journal} {Phys. Rev. A}\ }\textbf {\bibinfo {volume} {67}},\ \bibinfo {pages} {031401} (\bibinfo {year} {2003})}\BibitemShut {NoStop}%
\bibitem [{\citenamefont {Guo}\ \emph {et~al.}(2021)\citenamefont {Guo}, \citenamefont {Kroeze}, \citenamefont {Marsh}, \citenamefont {Gopalakrishnan}, \citenamefont {Keeling},\ and\ \citenamefont {Lev}}]{PhononOpticalLattice1}%
  \BibitemOpen
  \bibfield  {author} {\bibinfo {author} {\bibfnamefont {Y.}~\bibnamefont {Guo}}, \bibinfo {author} {\bibfnamefont {R.~M.}\ \bibnamefont {Kroeze}}, \bibinfo {author} {\bibfnamefont {B.~P.}\ \bibnamefont {Marsh}}, \bibinfo {author} {\bibfnamefont {S.}~\bibnamefont {Gopalakrishnan}}, \bibinfo {author} {\bibfnamefont {J.}~\bibnamefont {Keeling}}, \ and\ \bibinfo {author} {\bibfnamefont {B.~L.}\ \bibnamefont {Lev}},\ }\bibfield  {title} {\textit {\bibinfo {title} {An optical lattice with sound},\ }}\href {\doibase 10.1038/s41586-021-03945-x} {\bibfield  {journal} {\bibinfo  {journal} {Nature}\ }\textbf {\bibinfo {volume} {599}},\ \bibinfo {pages} {211–215} (\bibinfo {year} {2021})}\BibitemShut {NoStop}%
\bibitem [{\citenamefont {Wickenbrock}\ \emph {et~al.}(2012)\citenamefont {Wickenbrock}, \citenamefont {Holz}, \citenamefont {Wahab}, \citenamefont {Phoonthong}, \citenamefont {Cubero},\ and\ \citenamefont {Renzoni}}]{PhononOpticalLattice2}%
  \BibitemOpen
  \bibfield  {author} {\bibinfo {author} {\bibfnamefont {A.}~\bibnamefont {Wickenbrock}}, \bibinfo {author} {\bibfnamefont {P.~C.}\ \bibnamefont {Holz}}, \bibinfo {author} {\bibfnamefont {N.~A.~A.}\ \bibnamefont {Wahab}}, \bibinfo {author} {\bibfnamefont {P.}~\bibnamefont {Phoonthong}}, \bibinfo {author} {\bibfnamefont {D.}~\bibnamefont {Cubero}}, \ and\ \bibinfo {author} {\bibfnamefont {F.}~\bibnamefont {Renzoni}},\ }\bibfield  {title} {\textit {\bibinfo {title} {Vibrational Mechanics in an Optical Lattice: Controlling Transport via Potential Renormalization},\ }}\href {\doibase 10.1103/PhysRevLett.108.020603} {\bibfield  {journal} {\bibinfo  {journal} {Phys. Rev. Lett.}\ }\textbf {\bibinfo {volume} {108}},\ \bibinfo {pages} {020603} (\bibinfo {year} {2012})}\BibitemShut {NoStop}%
\bibitem [{\citenamefont {Jepsen}\ \emph {et~al.}(2020)\citenamefont {Jepsen}, \citenamefont {Amato-Grill}, \citenamefont {Dimitrova}, \citenamefont {Ho}, \citenamefont {Demler},\ and\ \citenamefont {Ketterle}}]{XXZ_chain1}%
  \BibitemOpen
  \bibfield  {author} {\bibinfo {author} {\bibfnamefont {P.~N.}\ \bibnamefont {Jepsen}}, \bibinfo {author} {\bibfnamefont {J.}~\bibnamefont {Amato-Grill}}, \bibinfo {author} {\bibfnamefont {I.}~\bibnamefont {Dimitrova}}, \bibinfo {author} {\bibfnamefont {W.~W.}\ \bibnamefont {Ho}}, \bibinfo {author} {\bibfnamefont {E.}~\bibnamefont {Demler}}, \ and\ \bibinfo {author} {\bibfnamefont {W.}~\bibnamefont {Ketterle}},\ }\bibfield  {title} {\textit {\bibinfo {title} {Spin transport in a tunable Heisenberg model realized with ultracold atoms},\ }}\href {\doibase 10.1038/s41586-020-3033-y} {\bibfield  {journal} {\bibinfo  {journal} {Nature}\ }\textbf {\bibinfo {volume} {588}},\ \bibinfo {pages} {403–407} (\bibinfo {year} {2020})}\BibitemShut {NoStop}%
\bibitem [{\citenamefont {Jepsen}\ \emph {et~al.}(2021)\citenamefont {Jepsen}, \citenamefont {Ho}, \citenamefont {Amato-Grill}, \citenamefont {Dimitrova}, \citenamefont {Demler},\ and\ \citenamefont {Ketterle}}]{XXZ_chain2}%
  \BibitemOpen
  \bibfield  {author} {\bibinfo {author} {\bibfnamefont {P.~N.}\ \bibnamefont {Jepsen}}, \bibinfo {author} {\bibfnamefont {W.~W.}\ \bibnamefont {Ho}}, \bibinfo {author} {\bibfnamefont {J.}~\bibnamefont {Amato-Grill}}, \bibinfo {author} {\bibfnamefont {I.}~\bibnamefont {Dimitrova}}, \bibinfo {author} {\bibfnamefont {E.}~\bibnamefont {Demler}}, \ and\ \bibinfo {author} {\bibfnamefont {W.}~\bibnamefont {Ketterle}},\ }\bibfield  {title} {\textit {\bibinfo {title} {Transverse Spin Dynamics in the Anisotropic Heisenberg Model Realized with Ultracold Atoms},\ }}\href {\doibase 10.1103/PhysRevX.11.041054} {\bibfield  {journal} {\bibinfo  {journal} {Phys. Rev. X}\ }\textbf {\bibinfo {volume} {11}},\ \bibinfo {pages} {041054} (\bibinfo {year} {2021})}\BibitemShut {NoStop}%
\bibitem [{\citenamefont {Brechtelsbauer}\ \emph {et~al.}(2025)\citenamefont {Brechtelsbauer}, \citenamefont {M\"ogerle},\ and\ \citenamefont {B\"uchler}}]{ColdAtomExperiment}%
  \BibitemOpen
  \bibfield  {author} {\bibinfo {author} {\bibfnamefont {K.}~\bibnamefont {Brechtelsbauer}}, \bibinfo {author} {\bibfnamefont {J.}~\bibnamefont {M\"ogerle}}, \ and\ \bibinfo {author} {\bibfnamefont {H.~P.}\ \bibnamefont {B\"uchler}},\ }\bibfield  {title} {\textit {\bibinfo {title} {Quantum simulation of spin-1 $XXZ$ Heisenberg models and the Haldane phase with dysprosium},\ }}\href {\doibase 10.1103/PhysRevA.111.032621} {\bibfield  {journal} {\bibinfo  {journal} {Phys. Rev. A}\ }\textbf {\bibinfo {volume} {111}},\ \bibinfo {pages} {032621} (\bibinfo {year} {2025})}\BibitemShut {NoStop}%
\bibitem [{\citenamefont {Yu}\ \emph {et~al.}(2024)\citenamefont {Yu}, \citenamefont {Guo}, \citenamefont {Liu}, \citenamefont {Zhang}, \citenamefont {Wang}, \citenamefont {Li},\ and\ \citenamefont {Yan}}]{RingLattice1}%
  \BibitemOpen
  \bibfield  {author} {\bibinfo {author} {\bibfnamefont {X.}~\bibnamefont {Yu}}, \bibinfo {author} {\bibfnamefont {Y.}~\bibnamefont {Guo}}, \bibinfo {author} {\bibfnamefont {Z.}~\bibnamefont {Liu}}, \bibinfo {author} {\bibfnamefont {Y.}~\bibnamefont {Zhang}}, \bibinfo {author} {\bibfnamefont {J.}~\bibnamefont {Wang}}, \bibinfo {author} {\bibfnamefont {J.}~\bibnamefont {Li}}, \ and\ \bibinfo {author} {\bibfnamefont {J.}~\bibnamefont {Yan}},\ }\bibfield  {title} {\textit {\bibinfo {title} {Generation of stable ultraviolet optical ring lattices using monolithic AlN metasurfaces for cooling atoms},\ }}\href {\doibase 10.1364/OME.520951} {\bibfield  {journal} {\bibinfo  {journal} {Opt. Mater. Express}\ }\textbf {\bibinfo {volume} {14}},\ \bibinfo {pages} {1201} (\bibinfo {year} {2024})}\BibitemShut {NoStop}%
\bibitem [{\citenamefont {Amico}\ \emph {et~al.}(2005)\citenamefont {Amico}, \citenamefont {Osterloh},\ and\ \citenamefont {Cataliotti}}]{RingLattice2}%
  \BibitemOpen
  \bibfield  {author} {\bibinfo {author} {\bibfnamefont {L.}~\bibnamefont {Amico}}, \bibinfo {author} {\bibfnamefont {A.}~\bibnamefont {Osterloh}}, \ and\ \bibinfo {author} {\bibfnamefont {F.}~\bibnamefont {Cataliotti}},\ }\bibfield  {title} {\textit {\bibinfo {title} {Quantum Many Particle Systems in Ring-Shaped Optical Lattices},\ }}\href {\doibase 10.1103/PhysRevLett.95.063201} {\bibfield  {journal} {\bibinfo  {journal} {Phys. Rev. Lett.}\ }\textbf {\bibinfo {volume} {95}},\ \bibinfo {pages} {063201} (\bibinfo {year} {2005})}\BibitemShut {NoStop}%
\bibitem [{\citenamefont {Ohlsson}\ and\ \citenamefont {Zhou}(2021)}]{LindbladToNH1}%
  \BibitemOpen
  \bibfield  {author} {\bibinfo {author} {\bibfnamefont {T.}~\bibnamefont {Ohlsson}}\ and\ \bibinfo {author} {\bibfnamefont {S.}~\bibnamefont {Zhou}},\ }\bibfield  {title} {\textit {\bibinfo {title} {Density-matrix formalism for $\mathcal{PT}$-symmetric non-Hermitian Hamiltonians with the Lindblad equation},\ }}\href {\doibase 10.1103/PhysRevA.103.022218} {\bibfield  {journal} {\bibinfo  {journal} {Phys. Rev. A}\ }\textbf {\bibinfo {volume} {103}},\ \bibinfo {pages} {022218} (\bibinfo {year} {2021})}\BibitemShut {NoStop}%
\bibitem [{\citenamefont {Monkman}\ and\ \citenamefont {Berciu}(2025)}]{LindbladToNH2}%
  \BibitemOpen
  \bibfield  {author} {\bibinfo {author} {\bibfnamefont {K.}~\bibnamefont {Monkman}}\ and\ \bibinfo {author} {\bibfnamefont {M.}~\bibnamefont {Berciu}},\ }\href {https://arxiv.org/abs/2411.14599} {\bibinfo {title} {Limits of the Lindblad and non-Hermitian description of open systems},\ } (\bibinfo {year} {2025}),\ \Eprint {http://arxiv.org/abs/2411.14599} {arXiv:2411.14599 [quant-ph]} \BibitemShut {NoStop}%
\bibitem [{\citenamefont {Roccati}\ \emph {et~al.}(2022)\citenamefont {Roccati}, \citenamefont {Palma}, \citenamefont {Ciccarello},\ and\ \citenamefont {Bagarello}}]{LindbladToNH3}%
  \BibitemOpen
  \bibfield  {author} {\bibinfo {author} {\bibfnamefont {F.}~\bibnamefont {Roccati}}, \bibinfo {author} {\bibfnamefont {G.~M.}\ \bibnamefont {Palma}}, \bibinfo {author} {\bibfnamefont {F.}~\bibnamefont {Ciccarello}}, \ and\ \bibinfo {author} {\bibfnamefont {F.}~\bibnamefont {Bagarello}},\ }\bibfield  {title} {\textit {\bibinfo {title} {Non-Hermitian Physics and Master Equations},\ }}\href {\doibase 10.1142/s1230161222500044} {\bibfield  {journal} {\bibinfo  {journal} {Open Systems and Information Dynamics}\ }\textbf {\bibinfo {volume} {29}} (\bibinfo {year} {2022}),\ 10.1142/s1230161222500044}\BibitemShut {NoStop}%
\bibitem [{\citenamefont {Romh\'anyi}\ \emph {et~al.}(2011)\citenamefont {Romh\'anyi}, \citenamefont {Totsuka},\ and\ \citenamefont {Penc}}]{BondOperator1}%
  \BibitemOpen
  \bibfield  {author} {\bibinfo {author} {\bibfnamefont {J.}~\bibnamefont {Romh\'anyi}}, \bibinfo {author} {\bibfnamefont {K.}~\bibnamefont {Totsuka}}, \ and\ \bibinfo {author} {\bibfnamefont {K.}~\bibnamefont {Penc}},\ }\bibfield  {title} {\textit {\bibinfo {title} {Effect of Dzyaloshinskii-Moriya interactions on the phase diagram and magnetic excitations of SrCu${}_{2}$(BO${}_{3}$)${}_{2}$},\ }}\href {\doibase 10.1103/PhysRevB.83.024413} {\bibfield  {journal} {\bibinfo  {journal} {Phys. Rev. B}\ }\textbf {\bibinfo {volume} {83}},\ \bibinfo {pages} {024413} (\bibinfo {year} {2011})}\BibitemShut {NoStop}%
\bibitem [{\citenamefont {Bhowmick}\ and\ \citenamefont {Sengupta}(2021)}]{BondOperator2}%
  \BibitemOpen
  \bibfield  {author} {\bibinfo {author} {\bibfnamefont {D.}~\bibnamefont {Bhowmick}}\ and\ \bibinfo {author} {\bibfnamefont {P.}~\bibnamefont {Sengupta}},\ }\bibfield  {title} {\textit {\bibinfo {title} {Weyl triplons in ${\mathrm{SrCu}}_{2}{({\mathrm{BO}}_{3})}_{2}$},\ }}\href {\doibase 10.1103/PhysRevB.104.085121} {\bibfield  {journal} {\bibinfo  {journal} {Phys. Rev. B}\ }\textbf {\bibinfo {volume} {104}},\ \bibinfo {pages} {085121} (\bibinfo {year} {2021})}\BibitemShut {NoStop}%
\end{thebibliography}
\end{document}